\documentclass{ws-ijmpa}
\usepackage[super,compress]{cite}
\usepackage{graphicx}
\begin{document}
\markboth{Greg Landsberg}
{Searches for extra spatial dimensions with the CMS detector at the LHC}

\catchline{}{}{}{}{}
%
%%%%%%%%%%%%%%%%%%%%%%%%%%%%%%%%%%%%%%%%%%%%%%%%%%%%%%%%%%%%%%%%%%%%
\def\MET{\mbox{${\hbox{$E$\kern-0.6em\lower-.1ex\hbox{/}}}_T$}} %missing ET

\title{SEARCHES FOR EXTRA SPATIAL DIMENSIONS WITH THE CMS DETECTOR AT THE LHC}

\author{GREG LANDSBERG}

\address{Brown University, Department of Physics, 182 Hope St, Providence, RI 02912, USA\\
landsberg@hep.brown.edu}

\maketitle

\begin{history}
Published 4 May 2015
\end{history}

\begin{abstract}
The success of the first three years of operations of the LHC at center-of-mass energies of 7 and 8 TeV radically changed the landscape of searches for new physics beyond the standard model and our very way of thinking about its possible origin and its hiding place. Among the paradigms of new physics that have been probed quite extensively at the LHC, are various models that predict the existence of extra spatial dimensions. In this review, the current status of searches for extra dimensions with the CMS detector is presented, along with prospects for future searches at the full energy of the LHC, expected to be reached in the next few years.

\end{abstract}

\ccode{PACS numbers: 04.50.Gh, 04.50.-h, 04.70.Dy, 06.30.Bp, 12.60.-i, 12.90.+b, 14.70.Kv, 14.80.Rt}

%\tableofcontents

\section{Introduction}	

Since the Large Hadron Collider (LHC) started its first successful high-energy operations in 2010, hopes that new physics beyond the standard model (SM) would appear any moment were running high. With the large accumulated amount of proton-proton data at center-of-mass energies $\sqrt{s} = 7$ and 8 TeV, more and more sophisticated searches for new physics came along. Among the theoretical paradigms tested to a great extent are the recent models with extra spatial dimensions, either flat or curved, that appeared about a decade ago and quickly gained a lot of attention from both theoretical and experimental communities. These models offer a different solution to the infamous ``hierarchy problem" that plagues the SM, and promise an exciting possibility of studying quantum gravity at the LHC, including the most mysterious of its manifestations: the black holes. 

In this review, the current status of searches for extra spatial dimensions and quantum gravity with the Compact Muon Solenoid (CMS) detector at the LHC is presented. While so far all these searches came empty-handed, several new experimental methods and techniques have been developed, which also apply to other searches for new physics. Among these techniques are an empirical method to predict QCD background in high-multiplicity events from low-multiplicity samples, as well as dedicated methods of reconstructing objects with high Lorentz boost, which may be the key signature for discovery of high-mass resonances predicted in a variety of beyond the SM (BSM) models, including the extra dimensional paradigm.

The negative results of these searches changed the very way we think about the ``naturalness" (i.e. non-fine-tuned solutions to the hierarchy problem) and paved the way to new strategies of looking for these phenomena, which will become possible once the design LHC energy is reached in the next few years.

\section{Setting the Scene}

Several models with extra dimensions have appeared in recent years as a follow-up to the original, nearly century-old idea of Kaluza and Klein (KK)\cite{KK1,KK2,KK3} to achieve a unification of electromagnetism and gravity by adding a compact fourth spatial dimension. While the KK model did not quite work, huge progress in string theory in the past quarter of century helped to revive the concept of extra dimensions and resulted in the modern attempts to utilize them to solve the hierarchy problem of the standard model, namely the very large ratio between the electroweak symmetry breaking (EWSB) and Planck energy scales.

Among several modern models with extra dimensions, which result in a rich LHC phenomenology, the following ones have been probed explicitly in the CMS experiment:
\begin{itemlist}
\item Model with large extra dimensions by Arkani-Hamed, Dimopoulos, and Dvali~\cite{add1,add2,add3} (ADD). In this model, only gravity propagates in $n$ flat dimensions assumed to be compactified with a radius $R$ either on a sphere or on a torus. All the SM particles are confined on a 3D membrane (``brane") in the (4+n)-dimensional space-time. Apparent weakness of gravity (i.e., the large value of the Planck mass, $M_P$) can then be explained due to the volume suppression of the fundamentally strong gravity in 3+n dimensions (fundamental Planck scale being $M_D \sim 1$ TeV), once one looks at its effects in 3 dimensions. Typical values of $n$ being considered are between 2 and 6, with the $n=2$ already being significantly constrained by direct gravitaty measurements at short distances and from astrophysical and cosmological observations.\cite{ADDreview1,ADDreview2,ADDreview3} The radius of extra dimensions depends on their number, and the relationship between $M_D$ and $M_P$ can be established via Gauss's law:\cite{add1} $M_P^2 = 8\pi M_D^{n+2} R^n$ (for extra dimensions compactified on a torus). Setting $M_D = 1$ TeV gives $R \sim 1$ mm ($n = 2$) to 1 fm ($n = 6$). These values are {\it large\/} compared to the characteristic scales of particle physics interactions, hence the name: {\it large extra dimensions\/}.
\item Randall-Sundrum (RS) model~\cite{RS1,RS2}  with a single, ``warped" extra dimension, which is realized in five-dimensional anti-deSitter space-time ($AdS_5$). The metric of the $AdS_5$ space is given by $ds^2 = \exp(-2kR|\varphi|)\eta_{\mu\nu}dx^\mu dx^\nu - R^2 d\varphi^2$, where $0 \le |\varphi| \le \pi$ is the coordinate along the compact dimension of radius $R$, $k$ is the curvature of the $AdS_5$ space, often referred to as the ``warp factor", $x^\mu$ are the conventional (3+1)-space-time coordinates, and $\eta_{\mu\nu}$ is the metric tensor of Minkowski space-time. Two 3-dimensional branes with equal and opposite tensions are positioned at the fixed points of the $S_1/Z_2$ orbifold in the $AdS_5$ space, at $\phi = 0$ (SM brane) and at $\phi = \pi$ (Planck brane). In this model, gravity is generated on the Planck brane, whereas at least some of the SM particles are confined to the SM brane, separated from the Planck brane in the extra dimension. Due to the warping of the extra dimension, operators with the characteristic size of $M_P$ on the Planck brane give rise to exponentially suppressed energy scales on the SM brane: $M_D = \bar{M_P}\exp(-\pi k R)$, where $\bar M_P \equiv M_P/\sqrt{8\pi}$ is the reduced Planck scale. Thus the EWSB scale can be connected to the Planck scale with a relatively low degree of fine tuning by requiring the product of the warp factor and the radius of the compact dimension: $kR \sim 10$. In this model $R$ could have a ``natural" value of $\sim 1/M_P$, thus offering a rigorous solution to the hierarchy problem. In certain variations of the RS model, some of the SM particles could be displaced w.r.t. the SM brane; therefore gravitational coupling may not be universal as it does depend on the geometrical overlap  of the graviton and SM particle wave functions in the extra dimension.
\item Universal extra dimensional model~\cite{UED} by Appelquist, Cheng, and Dobrescu. In this model, all SM particles can travel in $n$ extra dimensions and standard constraints from atomic physics and low-energy experiments are avoided by introducing a conserved quantum number, Kaluza-Klein parity $P_{\rm KK}$~\cite{KKP}, which prohibits SM particles ($P_{\rm KK} = 0$) to be directly coupled to an excited KK mode ($P_{\rm KK} > 0$). This quantum number acts similarly to the R-parity~\cite{RPV} in supersymmetric models and ensures that KK excitations can only be produced in pairs, thus significantly weakening many of potential constraints. Typical values of $n$ being considered are one and two.
\item The TeV$^{-1}$ extra dimensional model of Dienes, Dudas, and Gherghetta~\cite{TeV-1}. In this model, only gauge bosons propagate in an extra dimension with $R \sim 1$ TeV$^{-1}$, potentially achieving low-energy unification of the strong, electromagnetic, and weak forces.
\end{itemlist}

The phenomenology from the point of view of (3+1)-dimensional space-time in all these models can be represented by a tower of Kaluza-Klein (KK) excitations of particles propagating in extra dimensions. For compactified extra dimensions, such particle could only have quantized values of the momentum projection on the compact dimensions (cf. classical ``particle in a box" problem in quantum mechanics). From the point of view of a 3D observer, this quantized momentum in the direction orthogonal to the 3D brane appears as a ``tower" of massive states, with the $n$-th excitation having a mass of $m_n^2 = m_0^2 + (n/R)^2$, where $m_0$ is the mass of the zeroth KK mode, representing the ground state, or the particle confined to a 3D brane.

\section{Probing the ADD Model}

\begin{figure}[tbh]
\centerline{\includegraphics[width=0.95\textwidth]{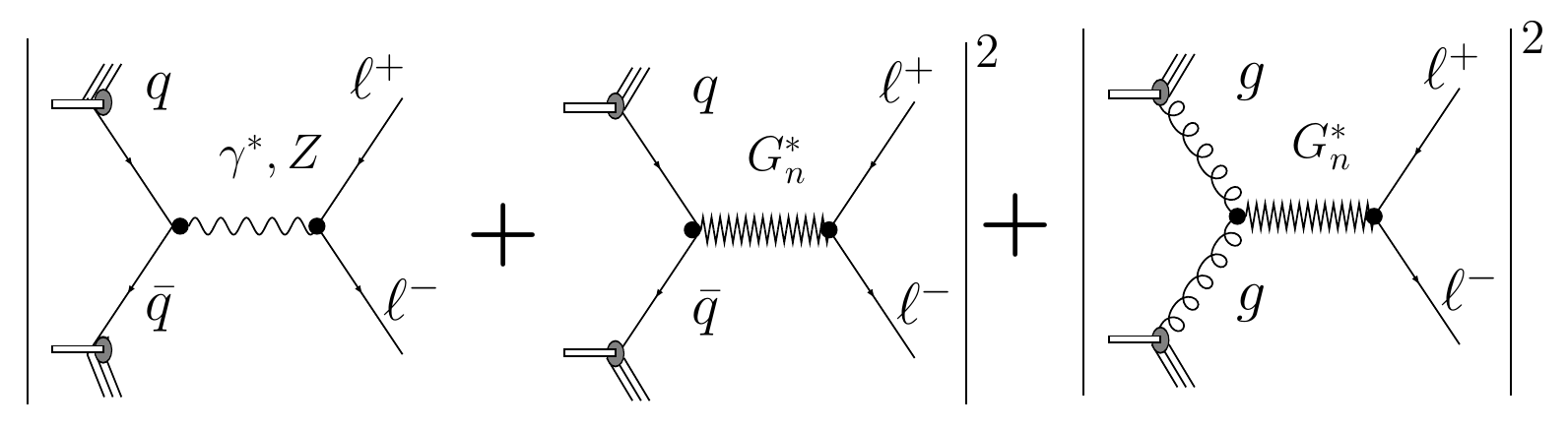}}
\caption{Leading-order diagrams contributing to Drell--Yan fermion pair production in the presence of low-scale gravity. From Ref. \protect\citen{Gupta}.\label{fig:DY}}
\end{figure}

There are three different classes of signatures that can be used to test the predictions of the ADD model:
\begin{itemlist}
\item {\it Virtual graviton effects.\/} In the ADD model, the KK modes of the graviton are very finely spaced in energy, as the size of extra dimensions is large by particle physics standards. Hence, it is experimentally impossible to distinguish between the individual KK states of the graviton, and their spectrum appears to be continuous. Each KK mode couples to the SM particles with the gravitational coupling, $\sqrt{4\pi G_N}$, where $G_N$ is the Newton's constant; since there are many KK modes that can be excited when the momentum transfer is large, the effective gravitational coupling becomes strong and the processes normally transmitted by other gauge bosons, e.g., Drell--Yan fermion pair production, can now be also carried via virtual gravitons (see Fig.~\ref{fig:DY}). Generally, the graviton-mediated diagrams interfere with their SM counterparts, resulting in a modification of the invariant mass spectrum of the final-state objects, particularly at large masses, for which the amplification of the gravitational interaction due to the  a large number of the accessible KK excitations becomes significant. This modification can be described via an effective field theory (EFT) approximation, similar to the compositeness operators~\cite{GRW,HLZ,Hewett}, where the ``compositeness scale" (i.e., the EFT cutoff) $M_S$, is expected to be of order $M_D$. The dependence on the number of extra dimensions is predicted to be weak.\cite{HLZ}

\item {\it Direct graviton emission.\/} Since gravity couples to energy-momentum tensor, any SM Feynman diagram vertex can be modified by attaching an extra graviton line to it. Thus, one can take any s-channel production vertex, e.g., $q\bar qg$ or $ggg$, and add a graviton to it, resulting in a $q\bar q \to gG$ or $gg \to gG$ process, where a gluon jet is recoiling against the graviton, which escapes in the extra dimensions. Just as in the virtual graviton case, when the momentum transfer is large, the corresponding effective coupling increases and the cross section for this reaction could become sizable. The signature for such a process in a collider detector is a single jet, countered by a significant missing momentum in the direction transverse to the beam, i.e. a {\it monojet}. This is a spectacular signature, quite different from usual SM jet pair production. Similarly, other objects can recoil against the graviton, giving rise to, e.g., monophoton signature. The probability of such a process is $\sim 1/M_D^{n+2}$ (see Ref. \citen{Peskin}), and thus strongly depends both on the fundamental Planck scale and on the number of extra dimensions.

\item {\it Black hole production.\/} One of the most spectacular signatures predicted~\cite{bf,dl,gt} within the ADD model is a possibility to produce microscopic black holes (BH) when the collision energy exceeds the value of the fundamental Planck scale $M_D$. These BHs may act either as their semi-classical counterparts in general relativity and evaporate via thermal Hawking radiation of various particle species; or they may have more complicated properties of their quantum precursors. Lacking the full picture of quantum gravity, it is impossible to have exact predictions on how these black holes evaporate, so many different possibilities should be considered. Since the Hawking temperature is inversely proportional to the Schwarzschild radius of such a BH~\cite{dl}, and the BH are tiny, the temperature is very hot and generally speaking the BH is expected to evaporate very fast either thermally (i.e., with an emission of a dozen or so particles, each carrying hundreds of GeV of energy), or via a decay, as a quantum state, into a pair of even more energetic particles. In either case, the signature is quite spectacular and can be also used to probe more general models with low-scale quantum gravity. The terminal stage of BH evaporation is not known. Some models predict that either a stable or unstable remnant with the mass of order of the fundamental Planck scale is formed as the evaporating BH mass approaches this Planck scale. In other models it is assumed that thermal evaporation continues to the very end.
\end{itemlist}

\subsection{Searches for virtual graviton effects}

Virtual graviton effects in CMS have been sought in dilepton, diphoton, and dijet data using the invariant mass spectrum of the two produced objects. In both cases, the interference with the corresponding SM processes is taken into account. The dominant background in the mass range of interest is the SM production of dileptons, diphotons, or dijets. This background, known to next-to-next-to-leading-order (dileptons) or next-to-leading-order (NLO, diphotons, dijets) precision is reliably estimated from simulation. No excesses were found in any of the channels. The most stringent limits on $M_S$ come from the dijet data at $\sqrt{s} = 8$ TeV with a sample corresponding to an integrated luminosity of 19.7 fb$^{-1}$, and reach 8 TeV at a 95\% confidence level (CL). The results of these searches are summarized in Table~\ref{table:virtual}. For the dilepton searches~\cite{EXO-12-061,EXO-12-046}, a NLO signal $K$-factor of 1.3 was assumed,\cite{K-factor-dilepton1,K-factor-dilepton2} while for the diphoton search~\cite{diphoton} a $K$-factor of 1.6 was used.\cite{K-factor-diphoton1,K-factor-diphoton2} For the dijet~\cite{dijet} search, leading-order (LO) signal cross-section was used. For the dependence of the limits on the number of extra dimensions, the convention of Ref.~\citen{HLZ} was used. 

\begin{table}[htb]
\tbl{Lower observed (expected) limits at a 95\% CL on the EFT cutoff $M_S$, in TeV, from various CMS searches.}
{\begin{tabular}{cccccc@{}cc}\toprule
$n$=2 & $n$=3 & $n$=4 & $n$=5 & $n$=6 & $n$=7 & Process & Data sample \\
\colrule
3.70 & 4.58 & 3.85 & 3.48 & 3.24 & 3.07 & \hphantom{00}$q\bar q \to \mu^+\mu^-$ & $\sqrt{s} = 8$ TeV\\
(3.69) & (4.57) & (3.84) & (3.47) & (3.23) & (3.06) &  & \hphantom{00}L = 20.6 fb$^{-1}$, Ref.~\citen{EXO-12-061}\\
\colrule
3.94 & 4.63 & 3.89 & 3.52 & 3.27 & 3.09 & \hphantom{00}$q\bar q \to e^+ e^-$ & $\sqrt{s} = 8$ TeV\\
(3.91) & (4.61) & (3.88) & (3.50) & (3.26) & (3.08) &  & \hphantom{00}L = 19.7 fb$^{-1}$, Ref.~\citen{EXO-12-061}\\
\colrule
4.35 & 4.94 & 4.15 & 3.75 & 3.49 & 3.30 & \hphantom{00}$q\bar q \to e^+ e^-$ & $\sqrt{s} = 8$ TeV\\
(4.37) & (4.93) & (4.14) & (3.74) & (3.48) & (3.30) &  \hphantom{00}or $\mu^+\mu^-$ & L = 19.7 -- 20.6 fb$^{-1}$, 
Ref.~\citen{EXO-12-061}\\
\colrule
         & 3.33 & 2.80 & 2.53 & 2.35 & 2.23 & \hphantom{00}$q\bar q \to \tau^+ \tau^-$ & $\sqrt{s} = 8$ TeV\\
 &  &  & &  &  &  & \hphantom{00}L = 19.7 fb$^{-1}$, Ref.~\citen{EXO-12-046}\\
\colrule
3.68 & 3.79 & 3.18 & 2.88 & 2.68 & 2.53 & \hphantom{00}$q\bar q \to \gamma\gamma$ & $\sqrt{s} = 7$ TeV\\
(3.77) & (3.85) & (3.24) & (2.93) & (2.73) & (2.58) &  & \hphantom{00}L = 2.2 fb$^{-1}$, Ref.~\citen{diphoton}\\
\colrule
6.9 & 8.4 & 7.1 & 6.4 & 5.9 &  & \hphantom{00}$pp \to jj$ & $\sqrt{s} = 8$ TeV\\
(6.6) & (8.0) & (6.8) & (6.1) & (5.7) &   &  & \hphantom{00}L = 19.7 fb$^{-1}$, Ref.~\citen{dijet}\\
\botrule
\end{tabular}
\label{table:virtual}}
\end{table}

Given that the values of $M_S$ and $M_D$, while of the same order of magnitude, can differ from one the other (and $M_S$ is generally expected to be lower than $M_D$), searches for virtual graviton effects (sensitive to $M_S$) are complementary to the ones for direct graviton emission (sensitive to $M_D$).

\subsection{Searches for direct graviton emission}

Direct graviton emission in CMS has been probed in the monophoton and monojet channels. 

In both cases, the dominant background comes from production of a $Z$ boson in association with a jet or a photon, with the $Z$ boson decaying invisibly in a pair of neutrinos. In the case of the monojet analysis, this background is estimated from control samples in data where the $Z$ boson decays into a pair of muons. The acceptance correction factor from leptonic to invisible decays comes from simulation. In the monophoton analysis, the background is estimated from simulation corrected for NLO effects and cross-checked with the $Z(\ell\ell)\gamma$ sample, where $\ell = e$ or $\mu$.

The second-dominant background originates from $W$+jet or $W$+photon production, with the $W$ boson decaying leptonically (including the $\tau\nu$ channel) and the lepton falling either below the selection threshold on the transverse momentum ($p_T$) or outside of the detector acceptance. In the case of the $\tau$-lepton decay, the events can pass the signal selections either because the $\tau$ lepton decays hadronically, yielding an extra soft jet in the final state, or leptonically, with the lepton falling below the minimum $p_T$ threshold. For the monojet analysis, this background is estimated from a control sample of $W(\mu\nu)$+jet events, with the acceptance correction factors derived from simulation. For the monophoton analysis, the estimate is based on simulation adjusted for the NLO $K$-factor.

In the case of monophotons, there is an additional sizable background from $W(e\nu)$ events with the electron misreconstructed as a photon. This probability is measured using control sample of $Z(ee)$ events and then applied to the $W(e\nu)$ control sample to estimate this background contribution to the signal sample.

\begin{table}[htb]
\tbl{Lower observed (expected) limits at a 95\% CL on the fundamental Planck scale  $M_D$, in TeV, 
from various CMS searches. Cross sections used to set limits are calculated at leading order.}
{\begin{tabular}{ccccc@{}cc}\toprule
$n$=2 & $n$=3 & $n$=4 & $n$=5 & $n$=6 & Process & Data sample \\
\colrule
 & 2.12 & 2.07 & 2.02 & 1.97 & \hphantom{00}$pp \to \gamma + \MET$ & $\sqrt{s} = 8$ TeV\\
 & (1.96) & (1.92) & (1.89) & (1.88) &  & \hphantom{00}L = 19.6 fb$^{-1}$, Ref.~\citen{EXO-12-047}\\
\colrule
 5.61 & 4.38 & 3.86 & 3.55 & 3.26 & \hphantom{00}$pp \to {\rm jet} + \MET$ & $\sqrt{s} = 8$ TeV\\
 (5.09) & (3.99) & (3.74) & (3.32) & (2.99) &  & \hphantom{00}L = 19.6 fb$^{-1}$, Ref.~\citen{EXO-12-048}\\
\botrule
\end{tabular}\label{table:direct}}
\end{table}

Data agree well with the SM predictions in both the monojet and monophoton channels, both at $\sqrt{s} = 7$ and 8 TeV. The most stringent limits on $M_D$ come from the monojet analysis of the $\sqrt{s} = 8$ TeV dataset corresponding to an integrated luminosity of 19.6 fb$^{-1}$ and reach 5 TeV at a 95\% CL. The limits are summarized in Table~\ref{table:direct}. In the monophoton case, LO cross section for direct graviton emission was used, as NLO effects have not been calculated for this process. For the monojets, both LO and NLO calculations are available. Only LO-based limits are shown in Table~\ref{table:direct} for easier comparison between the two channels. Since the $\sqrt{s} =8$ TeV limits are significantly more stringent than the 7 TeV ones, only the former are shown in the table. 

\begin{figure}[hbt]
\centerline{\includegraphics[width=0.51\textwidth]{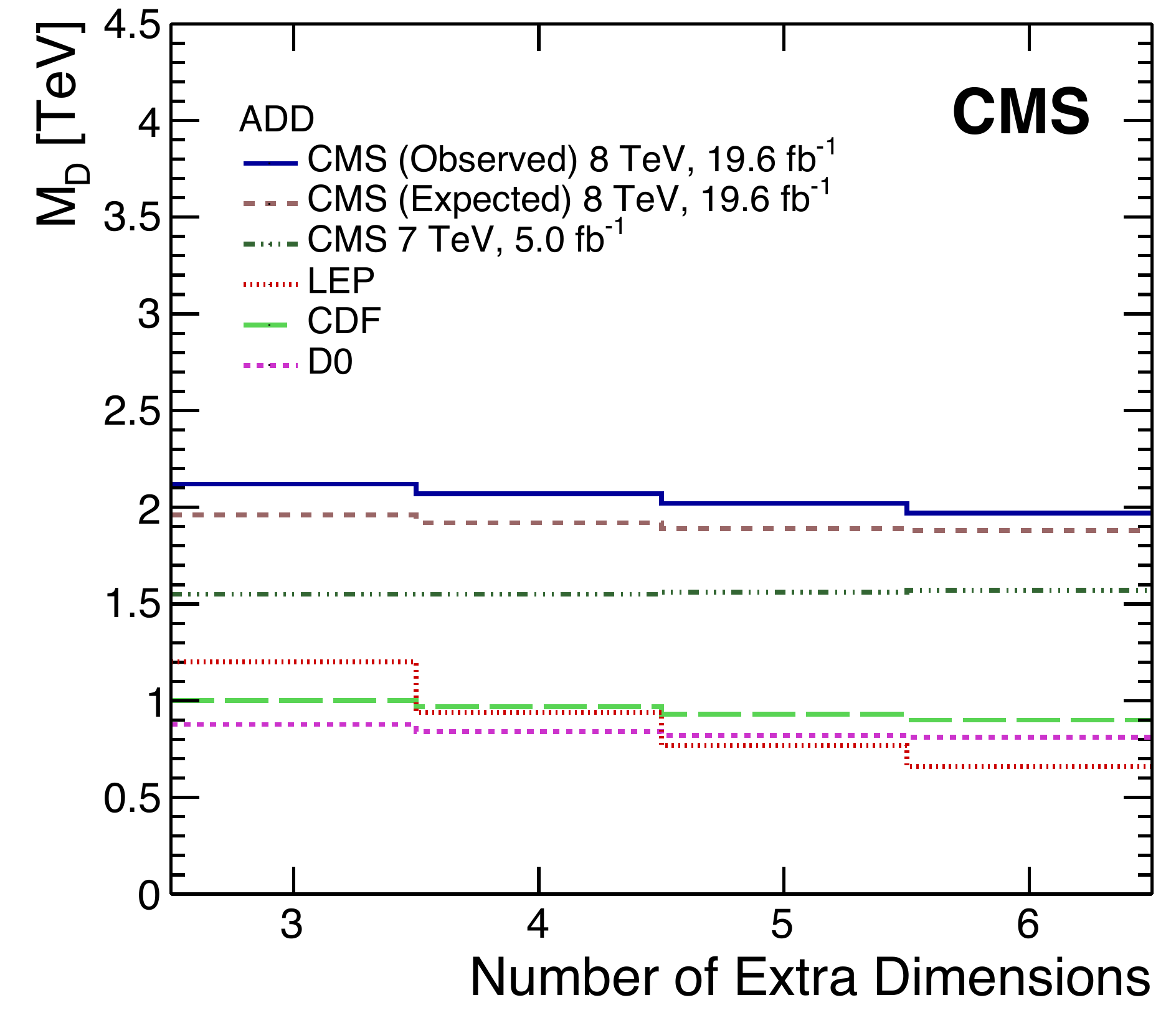}\
\includegraphics[width=0.47\textwidth]{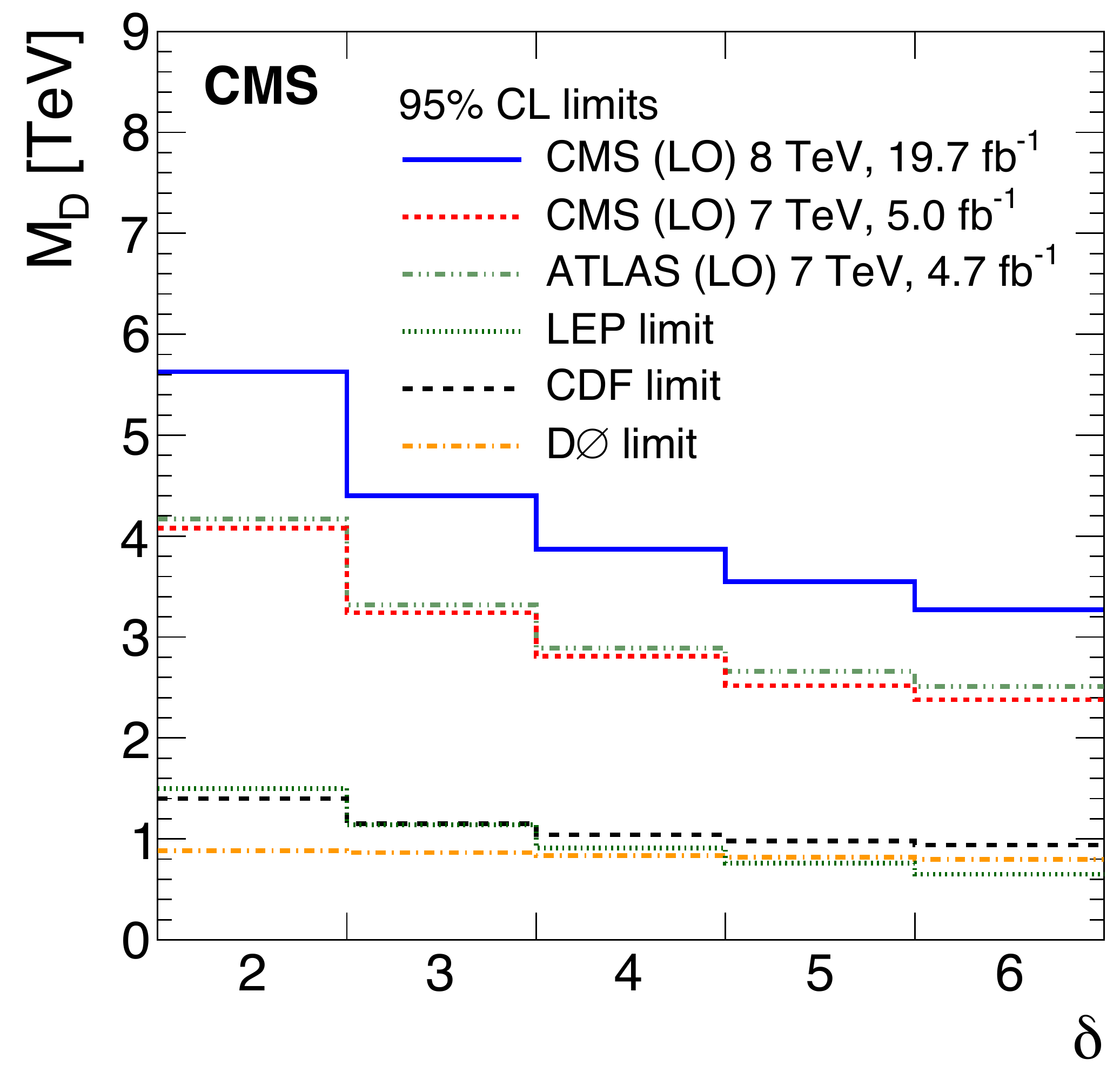}
}
\caption{Lower limits on the fundamental Planck scale $M_D$ at a 95\% CL from the monophoton (left) and monojet (right) analyses. From Refs \protect\citen{EXO-12-047,EXO-12-048}.
\label{fig:direct}}
\end{figure}

The limits from these two analyses, as well as their comparison with the $\sqrt{s} = 7$ TeV CMS results\cite{monophoton7,monojet7} and earlier measurements are shown in Fig.~\ref{fig:direct}.

\subsection{Limits on semiclassical black holes}

The CMS experiment has pioneered searches for microscopic black holes with the very first LHC data at $\sqrt{s} = 7$ TeV~\cite{CMSBH1}. Already with this small amount of data, corresponding to an integrated luminosity of just 0.035 fb$^{-1}$, stringent limits on semiclassical black hole production were set, using a novel method of predicting the dominant background coming from QCD multijet production. The analysis since then has evolved considerably and its later reincarnations\cite{CMSBH2,CMSBH3} probed many more additional BH models, including quantum black holes and string balls. Here we summarize the most recent analysis~\cite{CMSBH3} based on a $\sqrt{s} = 8$ TeV data sample with an integrated luminosity of 12.1 fb$^{-1}$.

This analysis was performed in inclusive final states, thus maximizing the sensitivity to BH production and decay. Semi-classical black holes are expected to evaporate in a large ($\sim 10$) number of energetic particles, emitted nearly isotropically, with the major fraction of the emitted particles being quarks and gluons~\cite{dl}, resulting in a multijet final state. Quantum effects and gray-body factors may change the relative fraction of emitted quarks and gluons, but generally it is expected that these particles still appear most often even in the decays of quantum black holes or their precursors, due to a large number of color degrees of freedom that quarks and gluons possess, compared to other SM particles.

The discriminating variable between the signal and the dominating QCD multijet background used in the search is the scalar sum of transverse momenta of all particles in the event, for which the $p_T$ value exceeds 50 GeV. This variable, $S_T$, was further corrected for any significant \MET\ in the event by adding the \MET\ value to the $S_T$ variable, if the former exceeds 50 GeV. The choice of $S_T$ as the discriminating variable is very robust and rather insensitive to the exact particle content in the process of BH evaporation, as well to the details of the final, sub-Planckian evaporation phase. The addition of \MET\ to the definition of $S_T$ further ensures high sensitivity of the search for the case of stable non-interacting remnant with the mass of order of the fundamental Planck scale, which may be produced in the terminal stage of the BH evaporation process.

The main challenge of the search is to describe the inclusive multijet background in a robust way, as the BH signal corresponds to a broad enhancement in $S_T$ distribution at high end, rather than a narrow peak. Since the BH signal is expected to correspond to high multiplicity of final-state particles, one has to reliably describe the background for large jet multiplicities, which is quite challenging theoretically, as higher-order calculations that fully describe multijet production simply do not exist. Thus, one can not rely on the Monte Carlo simulations to reproduce the $S_T$ spectrum correctly. 

\begin{figure}[hbt]
  \begin{center}
     \includegraphics[width=0.45\textwidth]{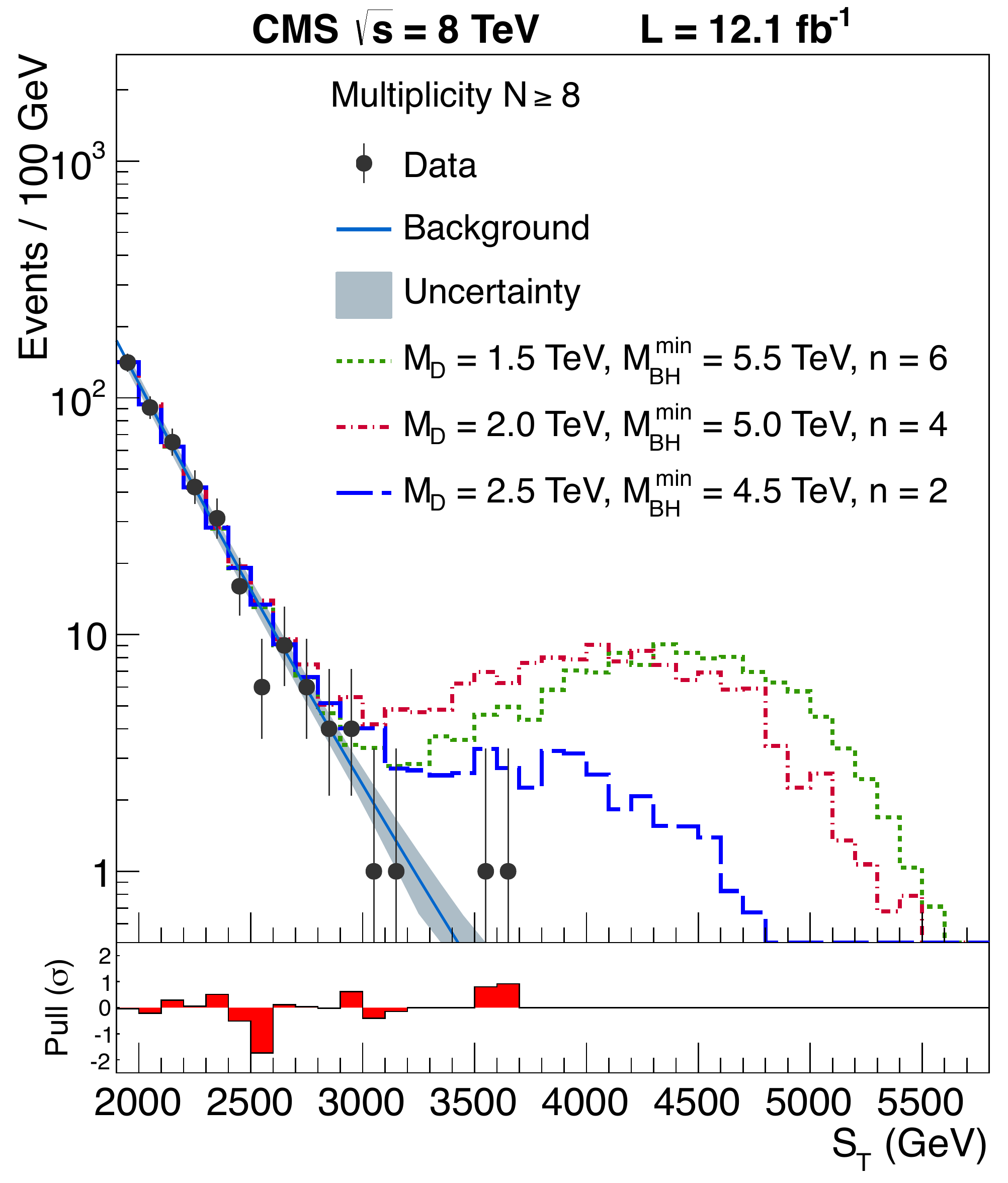}
     \includegraphics[width=0.54\textwidth]{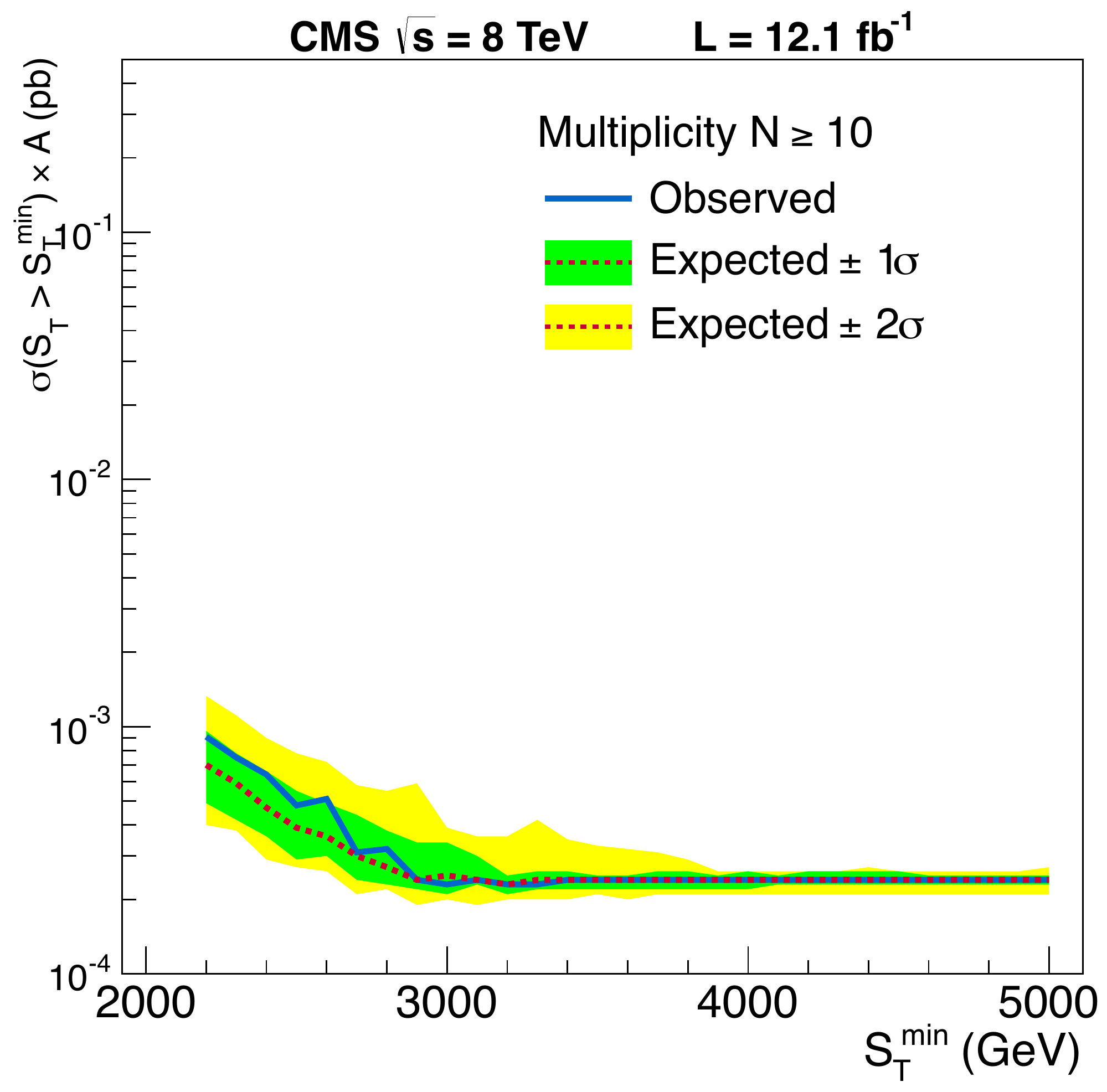}
  \end{center}
\caption{Left: predicted QCD multijet background with its uncertainties (the shaded band), data, and several reference BH signal benchmarks, as a function of $S_T$, in the final state with the multiplicity of eight or more particles. Right: model-independent upper limits at a 95\% CL on the cross section of a new physics signal decaying in the final state with 10 or more particles, as a function of the minimum $S_T$ requirement. From Ref. \protect{\citen{CMSBH3}}.}
\label{fig:CMSBH1}
\end{figure}

To overcome this problem, the CMS collaboration developed and utilized a novel method of predicting the QCD background directly from collision data. It has been found empirically, first via simulation-based studies, and then from the analysis of data at low jet multiplicities, that the shape of the $S_T$ distribution for the dominant QCD multijet background does not depend on the multiplicity of the final state, above the turn-on threshold. This observation, motivated by the way parton shower develops via nearly collinear emission, which conserves $S_T$, allows one to predict $S_T$ spectrum of a multijet final state using low-multiplicity QCD events, e.g. dijets or three-jet events. This provides a powerful method of predicting the dominant background for BH production by taking the $S_T$ shape from dijet events, for which the signal contamination is expected to be negligible, and normalizing it to the observed spectrum at high multiplicities at the low end of the $S_T$ distribution, where signal contamination is negligible even for large multiplicities of the final-state objects. The results are shown in Fig.~\ref{fig:CMSBH1} (left).

\begin{figure}[hbt]
  \begin{center}
    \includegraphics[width=0.49\textwidth]{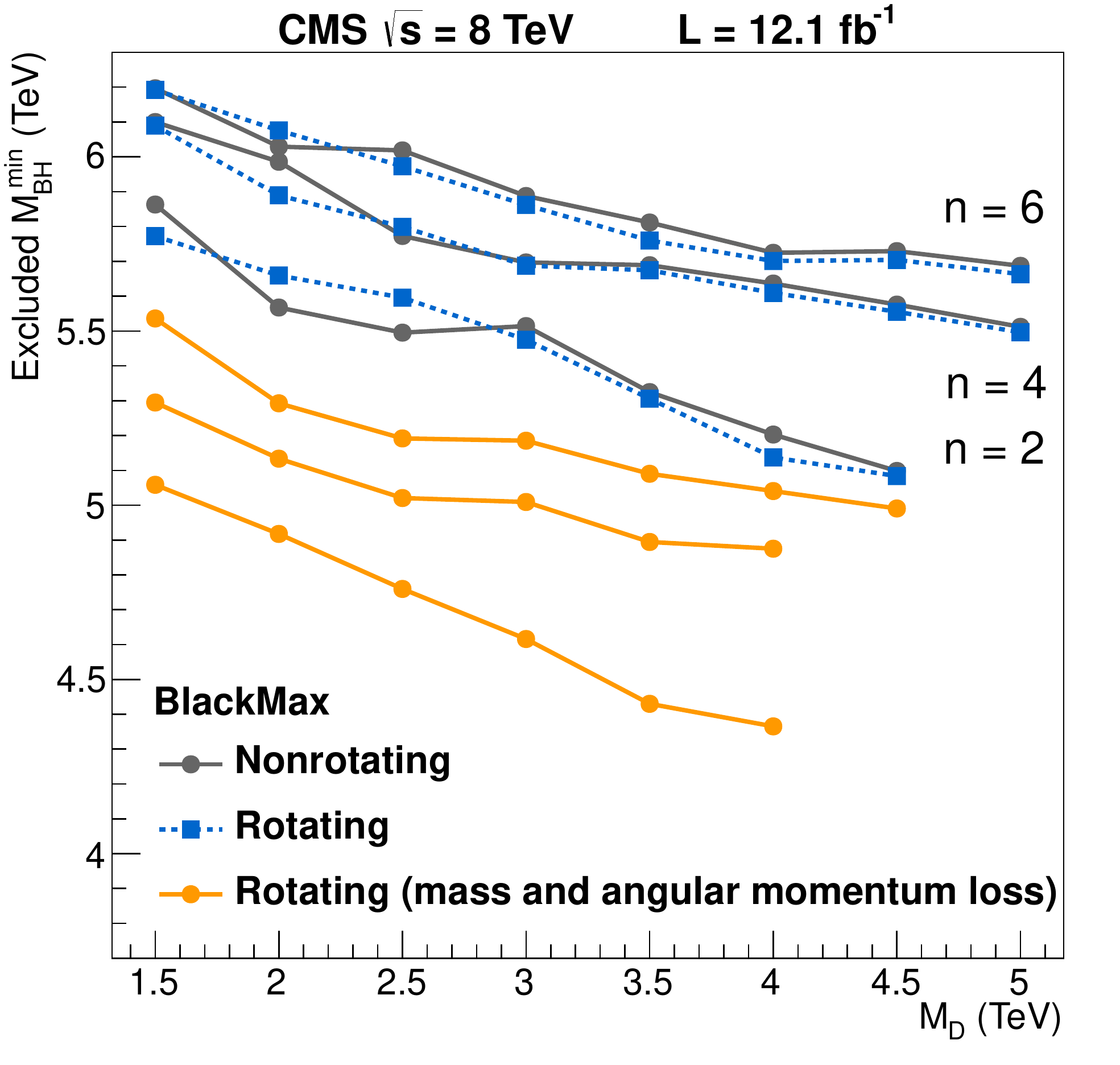}
    \includegraphics[width=0.49\textwidth]{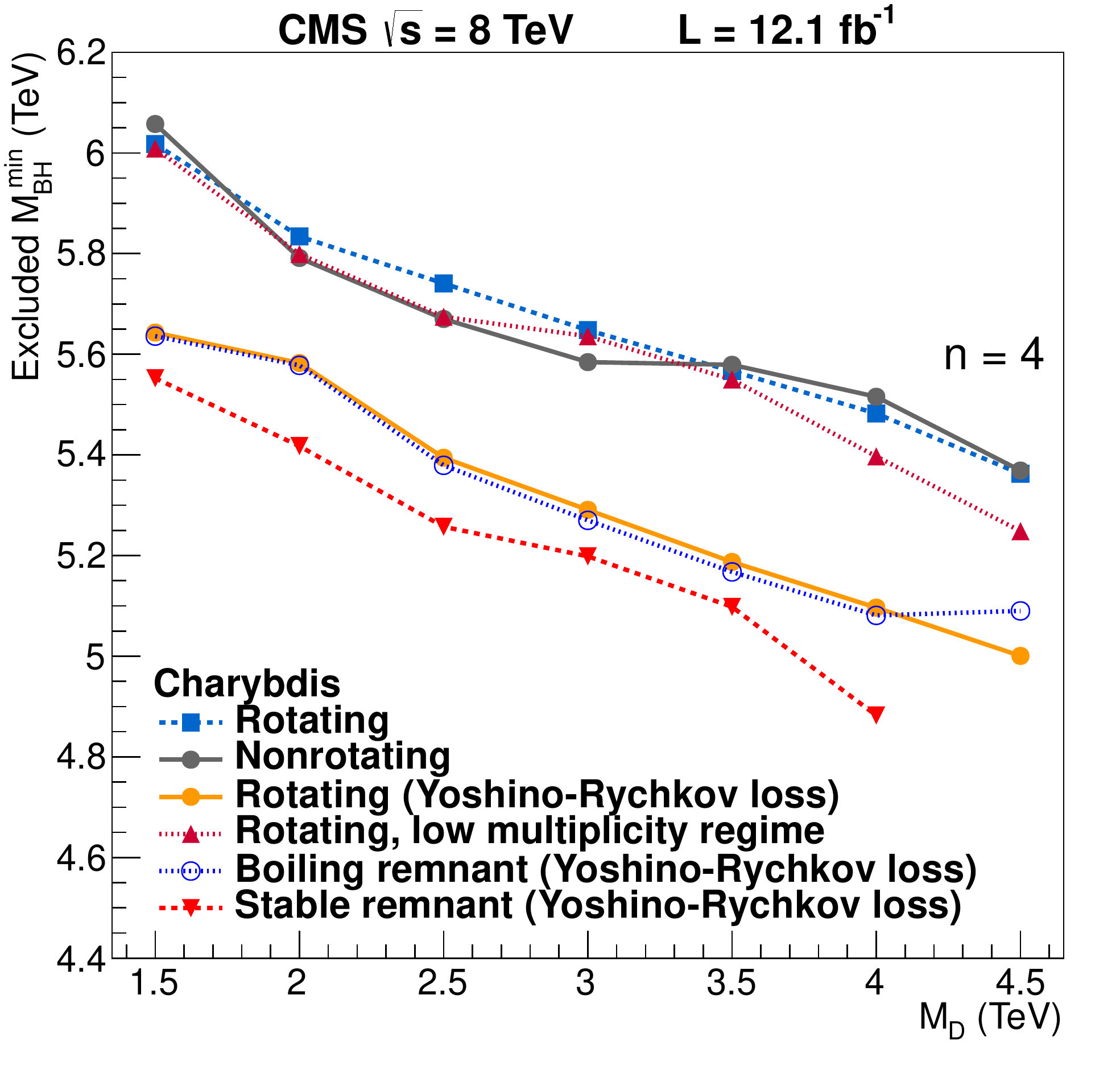}    
  \end{center}
\caption{Limits on the minimum BH mass as a function of the fundamental Planck scale for a few semi-classical benchmark models with and without BH rotation and with or without a remnant. Note that the semi-classical approximation used in setting these limits is not expected to hold for BH masses close to the Planck scale. From Ref. \protect{\citen{CMSBH3}}.}
\label{fig:CMSBH2}
\end{figure}

The CMS data with high final-state multiplicities is well fit by the background shape obtained from the dijet events, No excess characteristic of a BH production is observed even for highest multiplicities. This lack of an apparent signal can be interpreted in a model-independent way by providing a limit on the cross section for any new physics signal for $S_T$ values above a certain cutoff, for any given inclusive final state multiplicity. An example of such a limit is shown in Fig.~\ref{fig:CMSBH1} (right), for the final-state multiplicity $N \ge 10$. For signals corresponding to large values of $S_T$ (above 2 TeV or so) the cross section limit reaches $\sim 0.3$ fb. These limits can be compared with the production cross section for black holes in a variety of models and used to set limits on the minimum BH mass ($M_{\rm BH}$) that can be produced in these models. They are also applicable to other final states, e.g., from cascade decays of massive supersymmetric particles, thus making this search even more general. In fact, this very limit has been recently reinterpreted in terms of constraints on supersymmetry in Ref.~\citen{Strassler}.

For several specific models of semi-classical black holes the limits have been set explicitly via optimization of the analysis for these particular scenarios. These limits, although not very reliable for the black hole masses approaching the fundamental Planck scale, are shown in Fig.~\ref{fig:CMSBH2}.

As can be seen from these plots, the excluded minimum BH masses reach 5--6 TeV, which is close to the energy limit of the LHC machine at $\sqrt{s} = 8$ TeV. Therefore, the present searches are completely energy limited and will gain a significant boost once the LHC is restarted at $\sqrt{s} = 13$ TeV this year. Even with early high-energy data, significantly higher BH masses and fundamental Planck scale values could be probed. Nevertheless, it is clear that if an excess is found in this searches, we would be dealing with fundamentally quantum objects, given that the fundamental Planck scale limits from other searches already reached $\sim$ 5--8 TeV. Thus it is very important to also look for quantum black holes, which are expected to decay before they thermalize, resulting in just a few particles in the final state.

\subsection{Limits on quantum black holes and string balls}
\label{sec:QBH}

\begin{figure}[hbt]
  \begin{center}
    \includegraphics[width=0.45\textwidth]{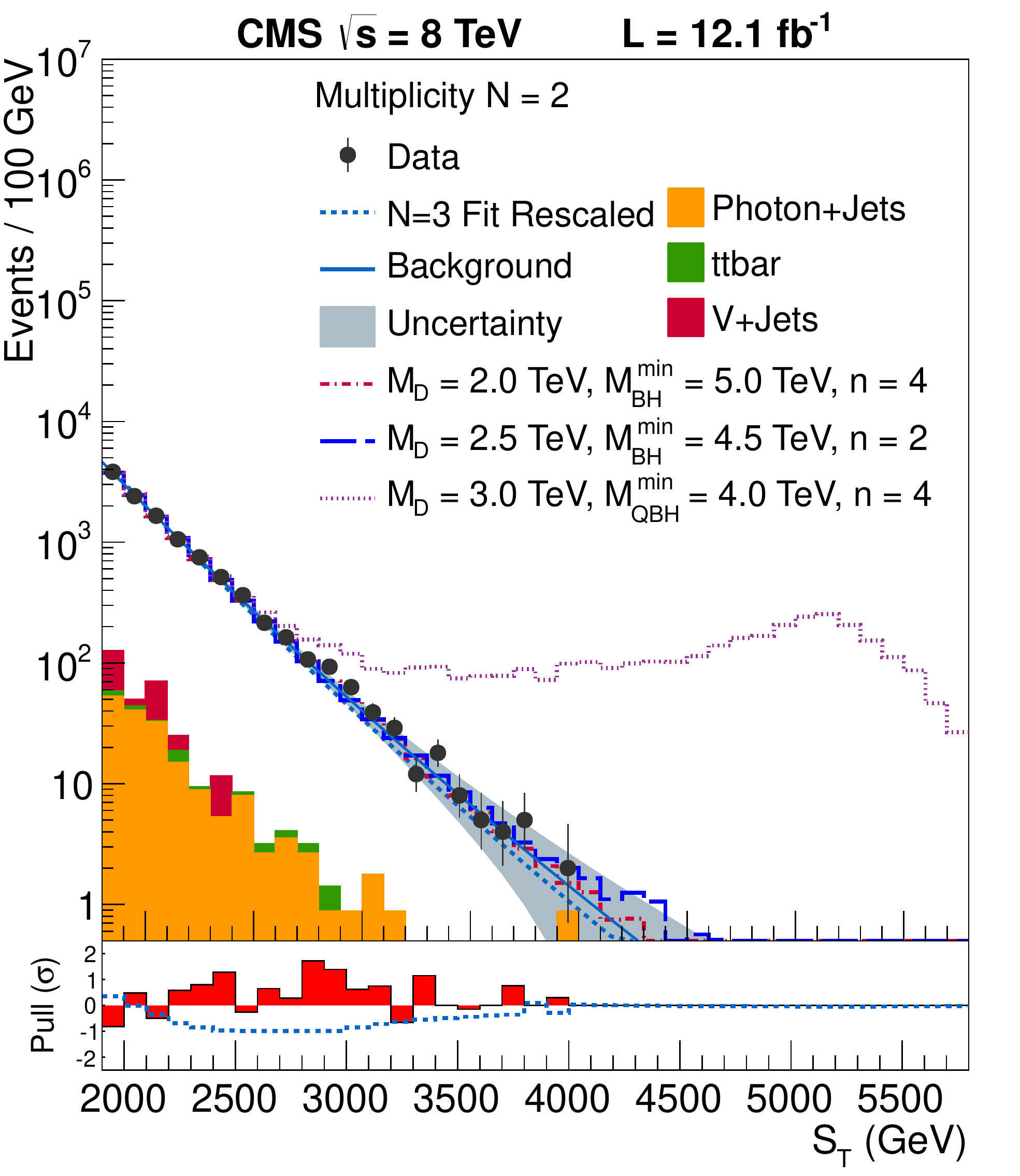}
    \includegraphics[width=0.54\textwidth]{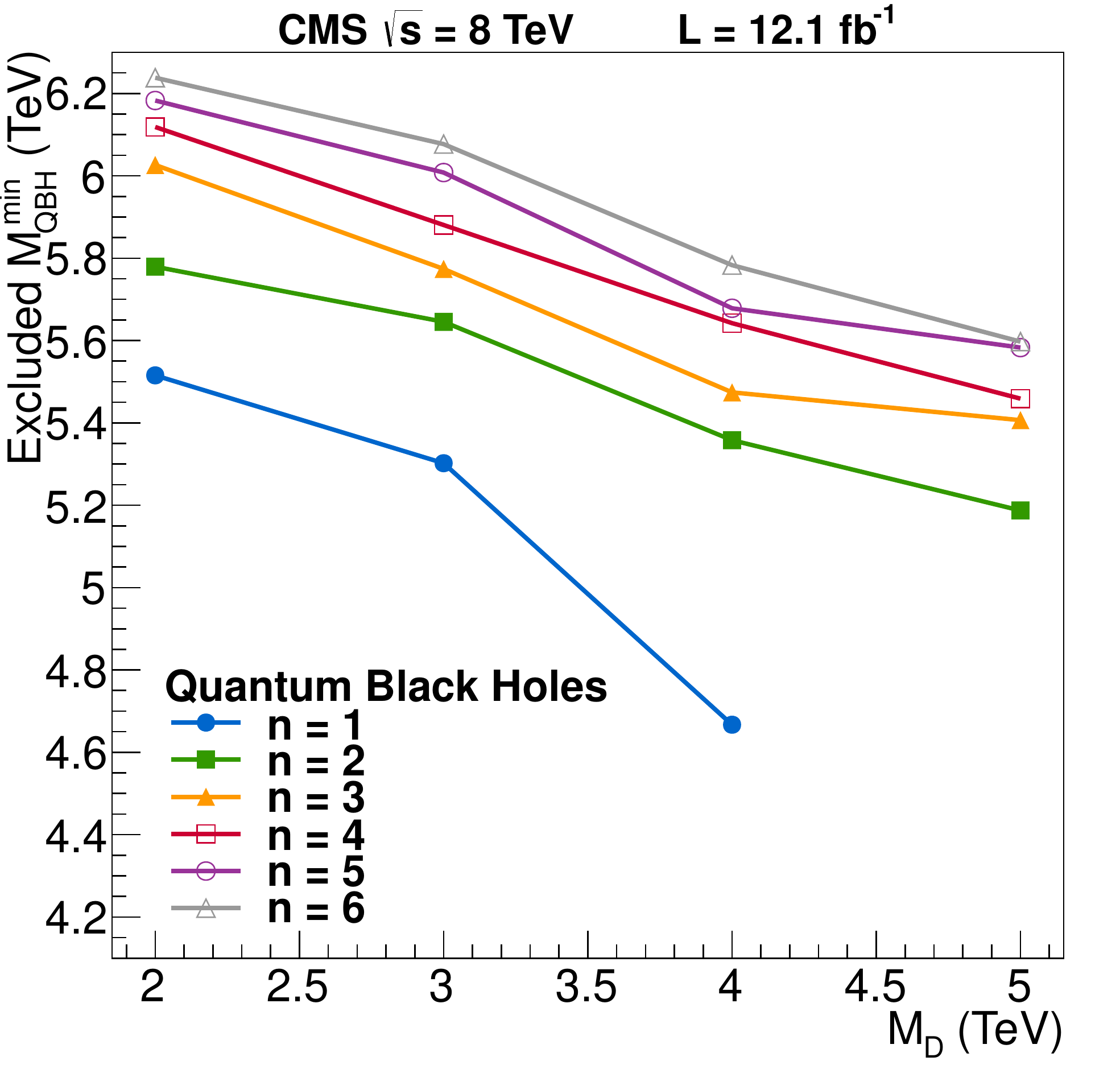}
  \end{center}
\caption{(left) The $N = 2$ $S_T$ spectrum with the background extrapolated from the low-$S_T$ region. The dashed blue line shows an alternative background prediction based on $N = 3$ $S_T$ spectrum. A benchmark quantum BH signal is shown with a dotted magenta line. (right) Limits on the minimum quantum BH mass for various numbers of extra dimensions. The $n = 1$ case corresponds to limits in the Randall-Sundrum model, in which the black hole formation is also possible (see Section~\protect\ref{sec:RSL}). From Ref. \protect{\citen{CMSBH3}}.}
\label{fig:CMSQBH}
\end{figure}

The same inclusive $S_T$-based analysis can be also used to set limits on quantum black holes. In this case, the result is based on the $S_T$ spectrum at low final-state object multiplicity ($N = 2$). Given that in this case one expects a contamination of the $S_T$ spectrum from signal, we predict QCD background at large $S_T$ by simply extrapolating the fit function from the low-$S_T$ range, which is used to determine the parameters of a smooth background fit and has negligible signal contamination (see Fig.~\ref{fig:CMSQBH}). We also check the background prediction obtained in this way by comparing it with the prediction based on the template obtained from the $N = 3$ spectrum (dashed blue line), which falls well within the uncertainties of the extrapolated background estimate.

\begin{figure}[hbt]
  \begin{center}
    \includegraphics[width=0.46\textwidth]{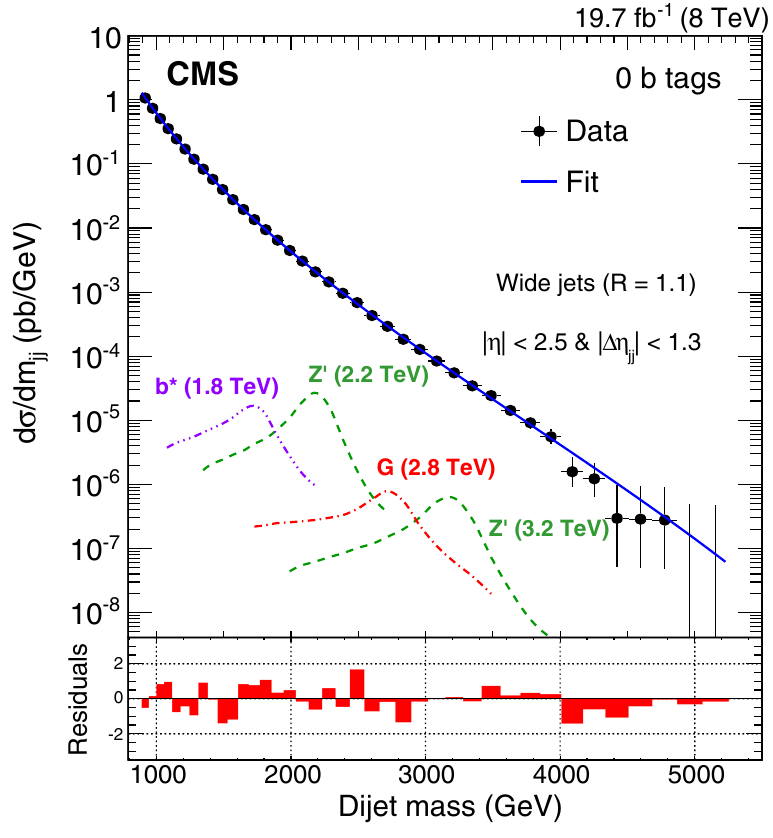}
    \includegraphics[width=0.52\textwidth]{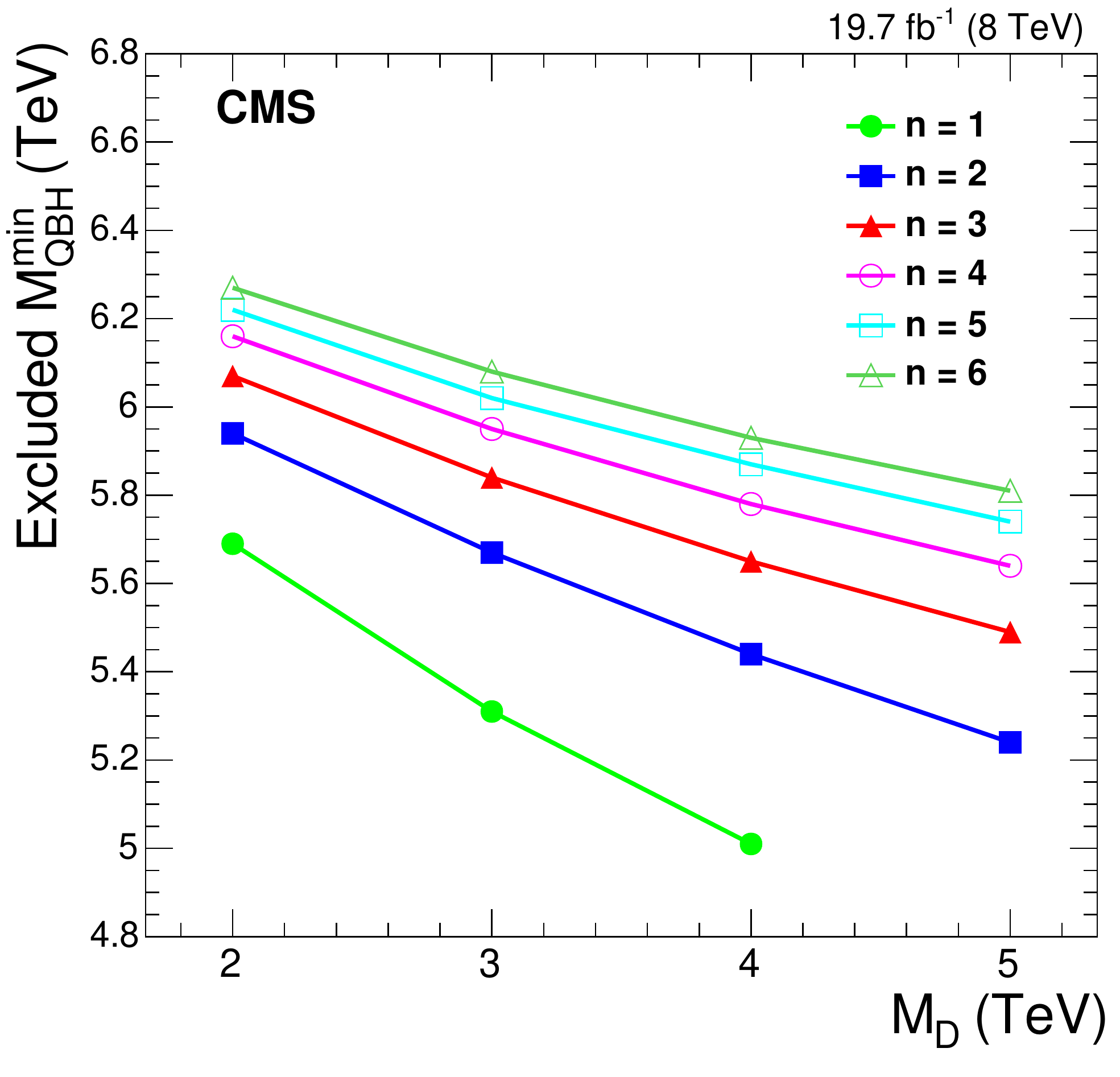}
  \end{center}
\caption{(left) The dijet invariant mass spectrum fitted with a smooth parametric background. (right) Lower limits on the quantum BH mass for various numbers of extra dimensions. The $n = 1$ case corresponds to limits in the Randall-Sundrum model (see Section~\protect\ref{sec:RSL}), in which the black hole formation is also possible. From Ref. \protect{\citen{CMSdijet8}}.}
\label{fig:CMSQBH1}
\end{figure}

Another search for quantum black holes in CMS was done using the dijet invariant mass spectrum\cite{CMSdijet8} obtained in the $\sqrt{s} = 8$ TeV dataset with an integrated luminosity of 19.7 fb$^{-1}$. In this case, the spectrum is fit with a sum of a smooth background template and a signal mass template obtained in simulation (see Fig.~\ref{fig:CMSQBH1} (left)). The analysis is a part of a general program of dijet resonance searches and set similar limits on quantum BH (see Fig.~\ref{fig:CMSQBH1} (right)) as the $S_T$-based analysis~\cite{CMSBH3} (see Fig.~\ref{fig:CMSQBH} (right)).

\begin{figure}[hbt]
  \begin{center}
    \includegraphics[width=0.7\textwidth]{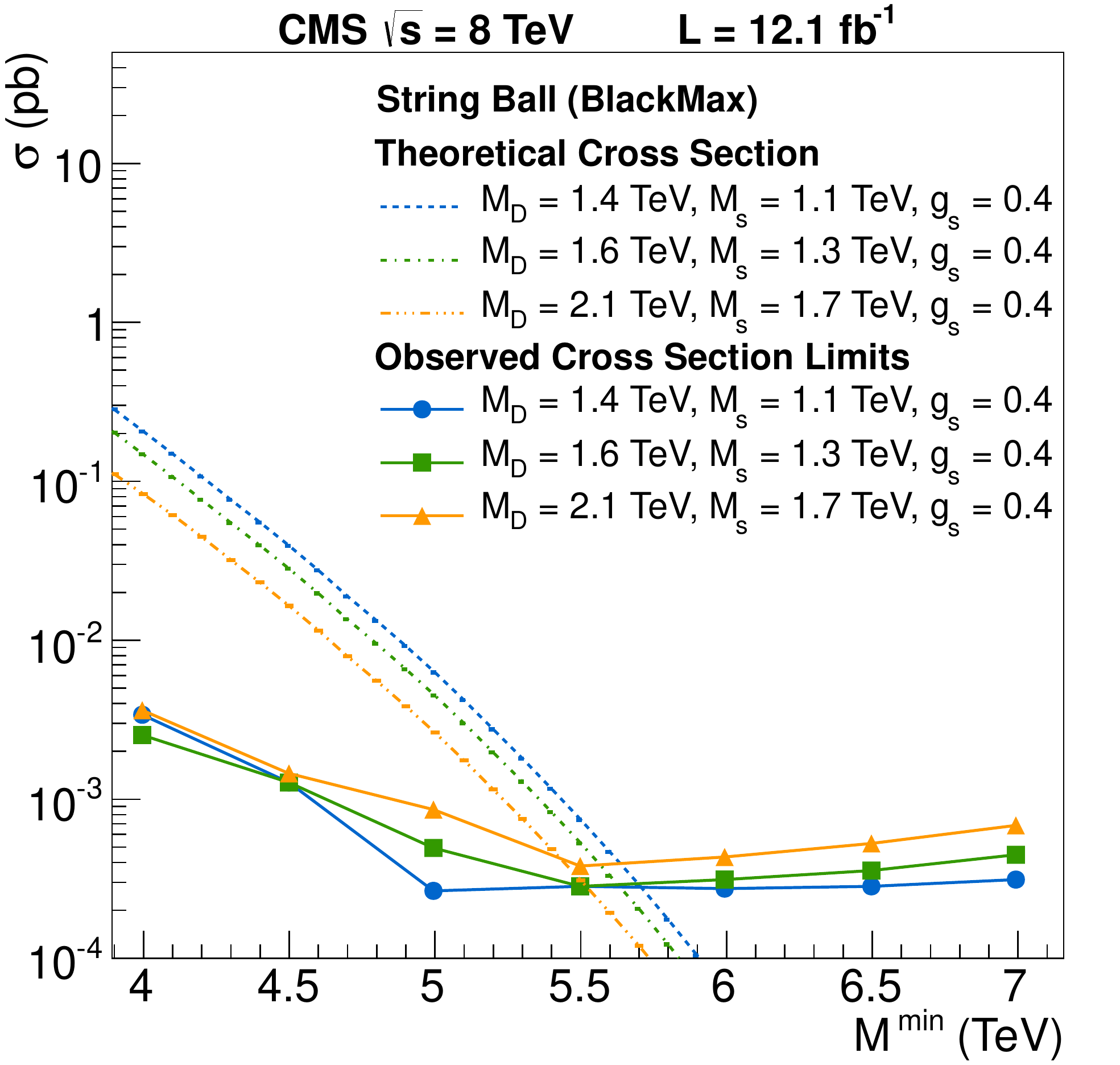}
  \end{center}
\caption{Lower limits on the minimum string ball mass for a range of values of the fundamental Planck scale $M_D$, string scale $M_s$, and string coupling constant $g_s$. From Ref. \protect{\citen{CMSBH3}}.}
\label{fig:sb}
\end{figure}

While the properties of quantum black holes remain an enigma, one could address a question of a light black hole formation in simple string theory models, which may correctly describe the effects of quantum gravity. One of the suggestions is that a precursor of a black hole is a highly excited excited string state, randomly folded in a ``string ball"~\cite{sb}. The properties of such a string ball are expected to be similar to those of a semi-classical black hole, with the exception that its evaporation takes place at a fixed, Hagedorn temperature~\cite{Hagedorn}, which does not depend on the string ball mass, but only on the string scale $M_s$ and string coupling constant $g_s$. Thus, the semi-classical black hole analysis~\cite{CMSBH3} can be also used to set limits on string balls, which are shown in Fig.~\ref{fig:sb} and reach 5.5--6.0 TeV.

\section{Probing the Randall-Sundrum Model}
\label{sec:RSL}

The LHC phenomenology of the RS model involves high-mass KK excitations of the graviton and, possibly, other bosons of the SM. There are many ``flavors" of the RS model, with various particles allowed to propagate in the bulk, so depending on a particular realization, the graviton may not couple to all the particle species in the same way, as was the case in the ADD model. Consequently, it's important to explore various decay channels of KK graviton and also look for KK excitations of other bosons.

Another possible experimental probe for RS model is production of microscopic BH~\cite{RSBH1,RSBH2,RSBH3}. In general these black holes are expected to act similar to the ADD black holes for the $n=1$ case, however, it has been argued that because of their higher temperature and shorter lifetime, RS black holes don't have time to thermalize and decay as quantum black holes in a small number of final-state particles. This signature has been covered in Section~\ref{sec:QBH} of this review..

While the RS model can be described using two parameters: the curvature of the anti-deSitter metric $k$ and the radius of the compact dimension $R$, phenomenologically, it's more convenient to use a different equivalent set of parameters, namely the mass of the first graviton KK excitation $M_1 \sim 1$ TeV and the dimensionless coupling $\tilde k \equiv k/ \bar M_P$. The spacing between the KK modes of the graviton is given by the subsequent zeroes $x_i$ of the Bessel function $J_1$ ($J_1(x_i) = 0$): $M_i = k x_i e^{-\pi k R} = \tilde{k} x_i M_D$. The coupling $\tilde k$ determines the strength of coupling of the graviton to the SM particles and the width of the KK excitations. Generally, it's expected that $\tilde k = O(10^{-2}) - O(10^{-1})$. Larger values of $\tilde k$ correspond to stronger coupling and broader KK resonances.

The CMS experiment has conducted a large variety of searches for KK excitations of the graviton and gluon in the RS model. The signatures probed so far include dileptons, pairs of quarks and gluons, including the top-quark pairs, and pairs of vector bosons. In the case of $t\bar t$, $WW$, and $ZZ$ final states, an additional complication arises from the fact that for large values of $M_1$, above $\sim 1$ TeV, the decay products (top quarks or vector bosons) are produced with high Lorentz boost, so their subsequent decay products are collimated and not always can be resolved, thus appearing as a single ``jet" in the detector. Special techniques, including jet substructure~\cite{JME-13-006,JME-13-007} are therefore utilized to maintain the sensitivity of these searches for large values of $M_1$. These jets of merged objects are referred to as ``jets" in the list of signatures below. Limits in the dilepton channels come from the analysis focused on a search for $Z'$ bosons\cite{EXO-12-061}; limits in the quark-antiquark and $gg$ final states come from general dijet resonance searches in the inclusive\cite{CMSdijet8} and $b\bar b$\cite{dijetbb} final states. Searches in the diboson and $t\bar t$ final state target the RS model specifically. The limits on RS gravitons from these CMS searches are summarized in Table~\ref{table:RS}.

\begin{table}[htb]
\tbl{Lower observed (expected) limits at a 95\% CL on the mass of the first KK graviton excitation $M_1$, in TeV, for two characteristic values of the coupling constant $\tilde k$. Signal cross sections used to set limits are calculated at NLO, except for the dijet searches, where a LO cross section was used.}
{\begin{tabular}{cc@{}cc}\toprule
 $\tilde k = 0.05$ &$\tilde k = 0.1$ & Process & Data sample \\
\colrule
  2.35 & 2.73 & \hphantom{00}$pp \to e^+e^- + \mu^+\mu^-$ & $\sqrt{s} = 8$ TeV \\
  & & \hphantom{00}L = 19.7 -- 20.6 fb$^{-1}$, Ref.~\citen{EXO-12-061} &  \\
\colrule
  & 1.58 & \hphantom{00}$pp \to q\bar q + gg$ (LO) & $\sqrt{s} = 8$ TeV\\
  & (1.43) & \hphantom{00}L = 19.6 fb$^{-1}$, Ref.~\citen{CMSdijet8} & \\
\colrule
 1.50 & 1.84 & \hphantom{00}$pp \to \gamma\gamma$ & $\sqrt{s} = 7$ TeV\\
  &  & \hphantom{00}L = 5.0 fb$^{-1}$, Ref.~\citen{diphoton} & \\
\colrule
  & 1.57 & \hphantom{00}$pp \to b\bar b$ (LO) & $\sqrt{s} = 8$ TeV\\
  &  & \hphantom{00}L = 19.6 fb$^{-1}$, Ref.~\citen{dijetbb} & \\
\colrule
  0.85 & 0.95 & \hphantom{00}$pp \to ZZ \to $ $eejj+\mu\mu jj$ & $\sqrt{s} = 7$ TeV\\
  &  & \hphantom{00}L = 4.9 fb$^{-1}$, Ref.~\citen{ZZlljj} & \\
\colrule
  0.88 & & \hphantom{00}$pp \to ZZ \to $ $ee+\mu\mu+\nu\nu$ + 1 ``jet" & $\sqrt{s} = 7$ TeV\\
   &  & \hphantom{00}L = 5.0 fb$^{-1}$, Ref.~\citen{arXiv:1211.5779} & \\
\colrule
  & 1.2 & \hphantom{00}$pp \to WW \to $ 2 ``jets"  & $\sqrt{s} = 8$ TeV\\
  & (1.3) & \hphantom{00}L = 19.7 fb$^{-1}$, Ref.~\citen{arXiv:1405.1994} & \\
\botrule
\end{tabular}\label{table:RS}}
\end{table}

\begin{figure}[hbt]
  \begin{center}
    \includegraphics[width=0.8\textwidth]{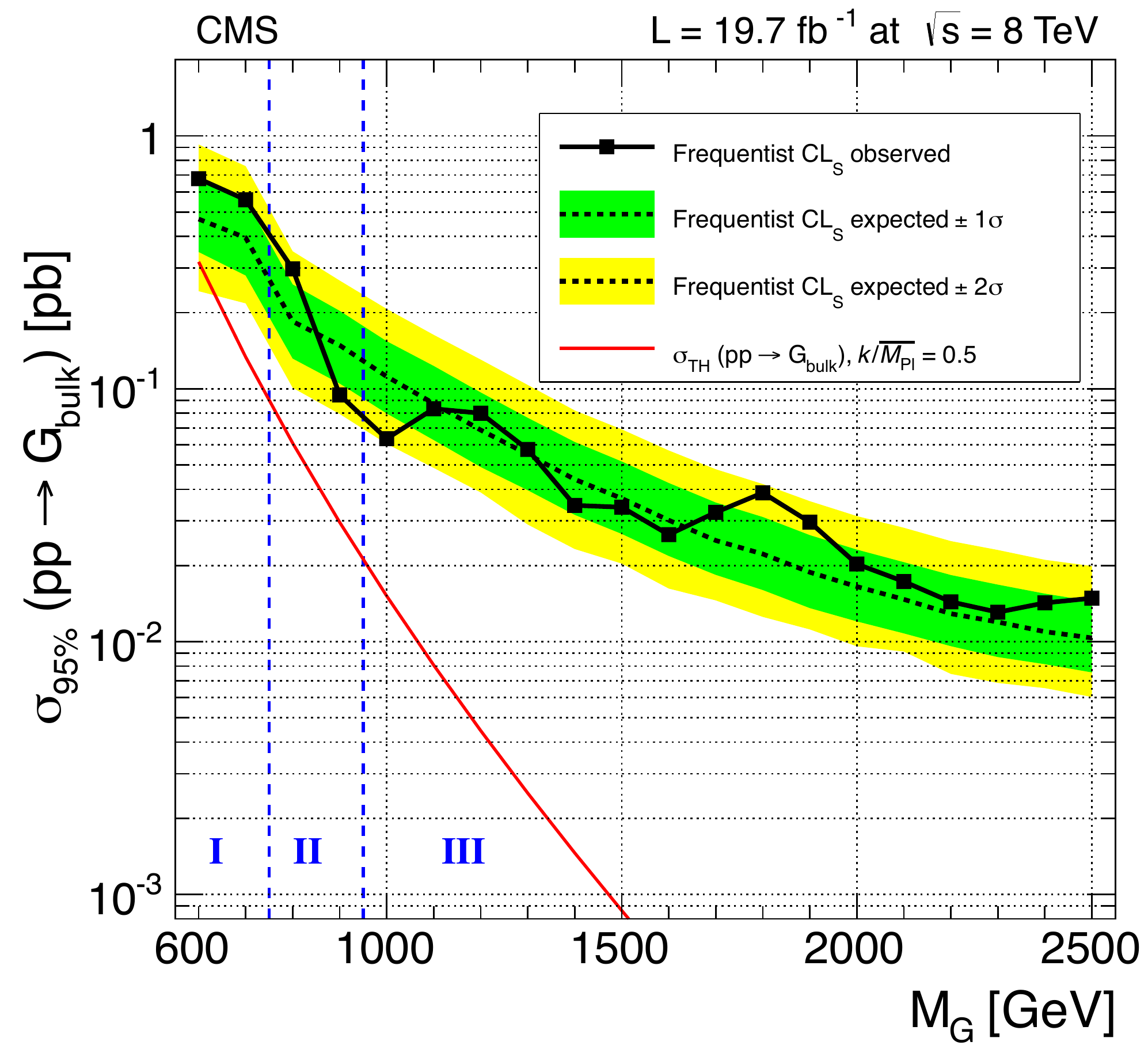}
  \end{center}
\caption{An observed (solid) and expected (dashed) 95\% CL upper limit on the bulk graviton production cross section obtained with the combination of semileptonic\protect{\cite{arXiv:1405.3447}} and all-hadronic\protect{\cite{arXiv:1405.1994}} analyses. The cross section for the production of a bulk graviton with $\tilde k=0.5$ is shown as a red solid curve. In region I, only the $\ell \ell$ + ``jet" channel contributes. In region II, both $\ell \ell$ + ``jet" (where $\ell = e$ or $\mu$) and $\ell \nu$ + ``jet" channels contribute. In region III, both the semi-leptonic and all-hadronic channels contribute.  From Ref. \protect{\citen{arXiv:1405.3447}}.}
\label{fig:bulk}
\end{figure}

The semileptonic\cite{arXiv:1405.3447} and all-hadronic\cite{arXiv:1405.1994} $WW/ZZ$ analyses also set limits on the bulk gravitons in a modified RS model~\cite{Agashe,Fitzpatrick,Antipin}. In this scenario, coupling of the graviton to light fermions and photons is
suppressed, while the production of gravitons via gluon fusion and their decays into a pair of massive gauge bosons is
sizable.The $\sqrt{s} = 8$ TeV analysis focuses mainly on the graviton mass range above 1 TeV, where the jet substructure techniques are more efficient than explicitly resolving two jets from the $Z$-boson decay; consequently it is not as sensitive at low masses. The analysis is not yet sensitive to the bulk gravitons with $\tilde k < 0.5$ in the entire mass range studied, so only the limits on the cross section are set (see Fig.~\ref{fig:bulk}). Note that CMS has conducted an earlier analysis at $\sqrt{s} = 7$ TeV  in the $ZZ \to eejj + \mu\mu jj$ final state\cite{ZZlljj} focused on the mass range below 1 TeV. That analysis claimed an exclusion limit of 0.645 TeV on the bulk graviton mass. However, the mass exclusion was based on the original theoretical calculations~\cite{Agashe}, which have been recently revised down by a factor of four~\cite{Carvalho}. The signal cross section curves in Fig.~\ref{fig:bulk} do reflect this updated cross section; hence this figure is not directly comparable with Fig. 3 (right) of Ref.~\citen{ZZlljj} in terms of the signal model.

\begin{figure}[hbt]
  \begin{center}
    \includegraphics[width=0.9\textwidth]{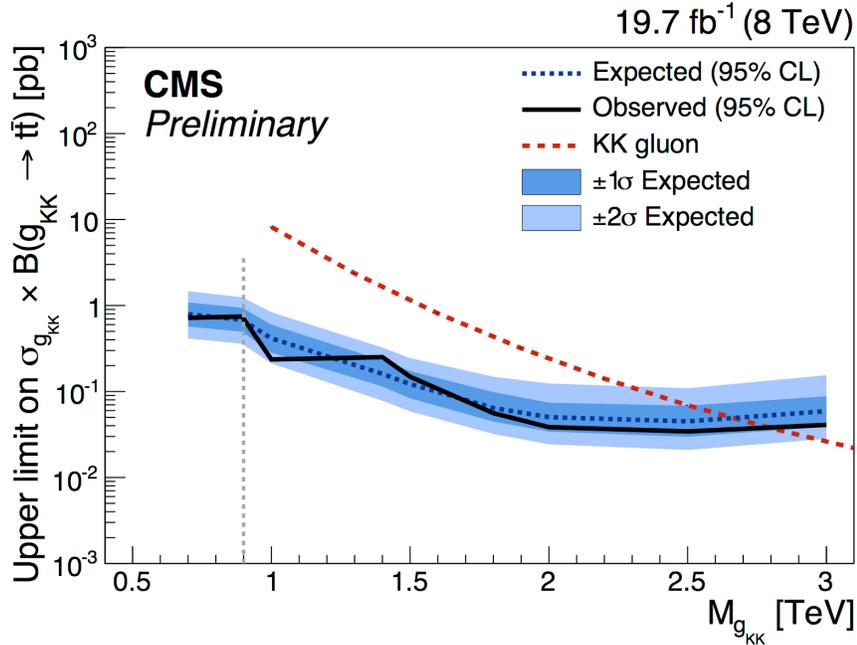}
  \end{center}
\caption{The 95\% CL upper limits on the product of the production cross section and the branching fraction of Kaluza--Klein excitation of a gluon, compared to the theoretical prediction. The $\pm 1\sigma$ and $\pm 2\sigma$ excursions from the expected limits are also shown. The vertical dashed line indicates the transition between the threshold semileptonic analysis \protect\cite{ttbar} and combined analysis, chosen based on their relative sensitivity. From Ref. \protect{\citen{ttbar-new}}.}
\label{fig:gKK}
\end{figure}

A search for a KK excitation of a gluon in the RS model with gluons in the bulk~\cite{AgashegKK} has been done in the $t\bar t$ decay mode, both in semileponic (resolved and boosted topologies) and all-hadronic decay modes of the $t\bar t$ quark pair. The combined analysis\cite{ttbar} uses jet substructure techniques to identified a jet originating from a collimated decay $t \to Wb \to jjb \to$ ``jet", by looking for an object consistent with a hadronic decay of the $W$ boson present within the jet, as well as considering the jet mass value. The analysis set a lower mass limit of 2.5 TeV (2.4 TeV expected) at a 95\% CL. Recently, this result has been superseded by a more advanced jet substructure analysis in the all-hadronic channel, and the dilepton channel has been also analyzed\cite{ttbar-new}. The combination of all there channels improved the earlier limit to 2.8 TeV (2.7 TeV expected) at a 95\% CL (see Fig.~\ref{fig:gKK}), which is the most stringent limit on KK gluons to date.

\section{Searches for Universal Extra Dimensions}

\begin{figure}[hbt]
  \begin{center}
    \includegraphics[width=0.7\textwidth]{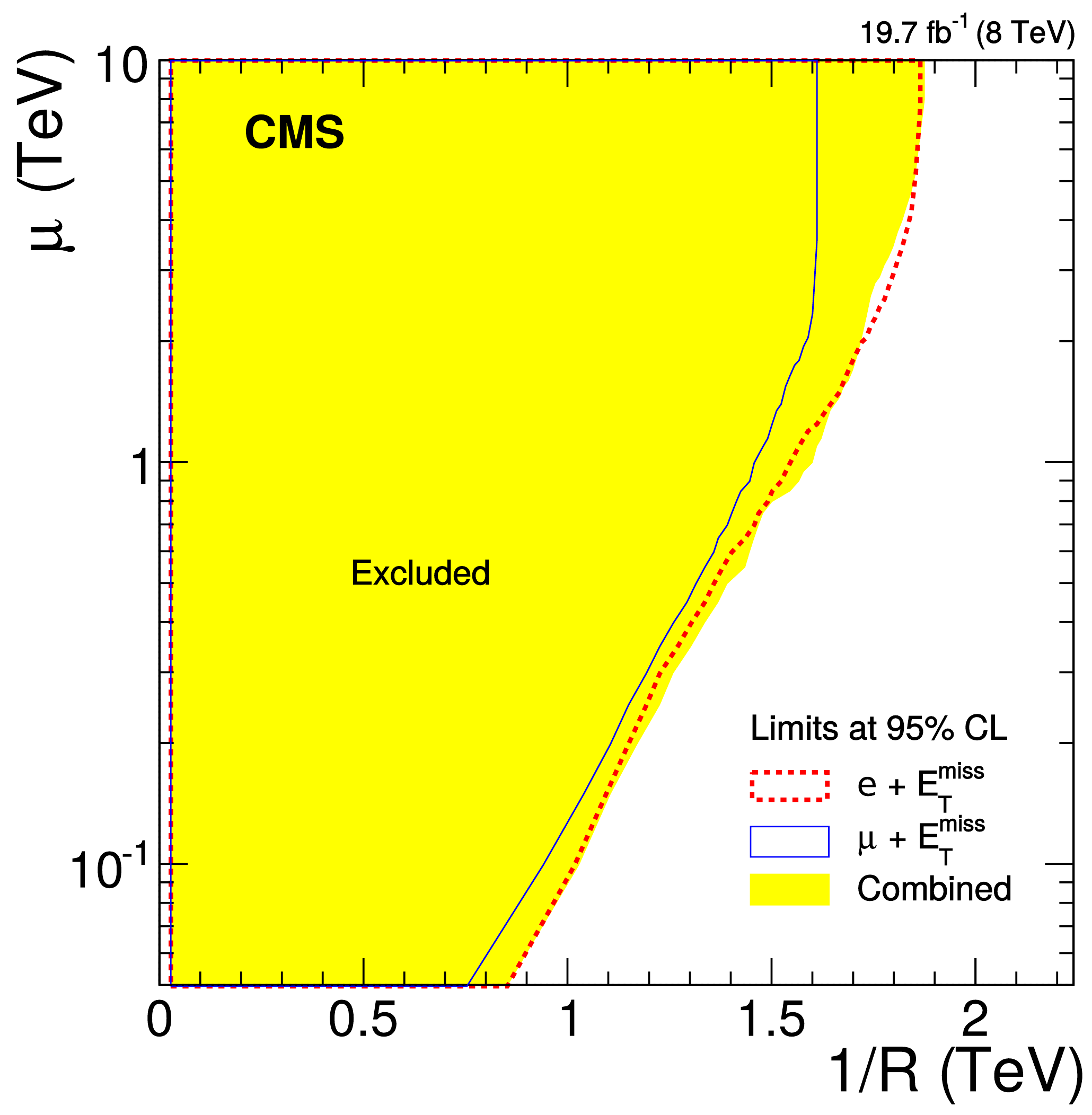}
  \end{center}
\caption{The 95\% confidence limits on the split-UED model parameters $\mu$ and $1/R$ derived from the $W'$ mass limits taking into account the corresponding width of the $W^{(2)}_{\rm KK}$. The limit for the electron (red dotted line) and the muon channel (blue line) individually along with their combination (yellow). From Ref. \protect{\citen{EXO-12-060}}.}
\label{fig:UED}
\end{figure}

Searches for universal extra dimensions in CMS have been conducted so far in a few specific UED models. One of them,\cite{splitUED,splitUED1,splitUED2} with a single extra dimension and bulk fermions, often referred to as ``split-UED model". In this model, only even KK excitations of gauge bosons, e.g., of the $W$ boson, can couple to SM fermions, due to KK parity conservation. Thus, one could search for the decay of the second KK excitation of the $W$ boson, W$^{(2)}_{KK}$, into a lepton (e or $\mu$) and a neutrino. This signature is also used to search for additional $W$-like gauge bosons, $W'$. The most stringent limits to date come from the CMS search\cite{EXO-12-060} based on the $\sqrt{s} = 8$ TeV sample corresponding to an integrated luminosity of 20 fb$^{-1}$. The limit is set on the inverse size of the compact extra dimension, $1/R$, as a function of the bulk mass parameter of 5D fermion fields $\mu$.\cite{splitUED1,splitUED2}. These limits are shown in Fig.~\ref{fig:UED} and reach $1/R$ of nearly 2 TeV for large values of $\mu$.

Another UED model probed at CMS is the one where the UED space is embedded in a space with additional large extra dimensions, where only gravity propagates. The lightest KK particle in this model is the first KK excitation of the photon, $\gamma^{(1)}_{KK}$, which then decays into a photon and a graviton. The resulting signature for the pair-produced $\gamma^{(1)}_{KK}$ particles (either directly or via cascade decay of other KK particles) is two photons and \MET\ due to undetected gravitons. This is a similar final state as expected in supersymmetric scenarios with gauge-mediated supersymmetry breaking (GMSB). A CMS GMSB supersymmetry search\cite{GMSB} at $\sqrt{s} = 7$ TeV based on a data sample with an integrated luminosity of 4.69 fb$^{-1}$ has been also interpreted in terms of limits on this UED scenario. The number of large extra dimensions $n$ in this scenario was varied between 2 and 6. For $n \ge 3$ decays involving a heavy graviton with mass of order $1/R$ dominate, while
for $n = 2$ decays involving massless gravitons prevail.\cite{Macesanu} The observed data in this analysis agree with the SM predictions, which allowed to set a limit of  $1/R > 1.38$ TeV for $n = 6$ at a 95\% CL, with an expected limit of 1.35 TeV.  For $n = 2$ the observed (expected) lower limit is reduced to 1.35 (1.34) TeV. These are the most stringent limits on the probed UED model to date.

\section{Searches for TeV$^{-1}$ Extra Dimensions}

In the model\cite{TeV-1} with extra dimensions with a size $R \sim$ TeV$^{-1}$, there is no KK parity conservation, so one could look directly for production of the first excited states of the $W$ or $Z$ bosons. The search is similar to the standard searches for $W'$ and $Z'$ bosons with the SM-like couplings, with the caveat that there is strong negative interference between the non-resonant $W^*$ or Drell--Yan production and the first KK excitation at approximately half the resonance mass. This effect needs to be accounted for in the limit-setting procedure.

While the $W'$ search~\cite{EXO-12-060} has been reinterpreted in terms of TeV$^{-1}$ extra dimensions, the current limit from direct searches are still significantly weaker than the indirect limits from LEP (see, e.g., Ref. \citen{Cheung}), so they are still more in a realm of the next LHC run. 

\section{Other Searches for Low-Scale Quantum Gravity}

\begin{figure}[hbt]
  \begin{center}
    \includegraphics[width=0.7\textwidth]{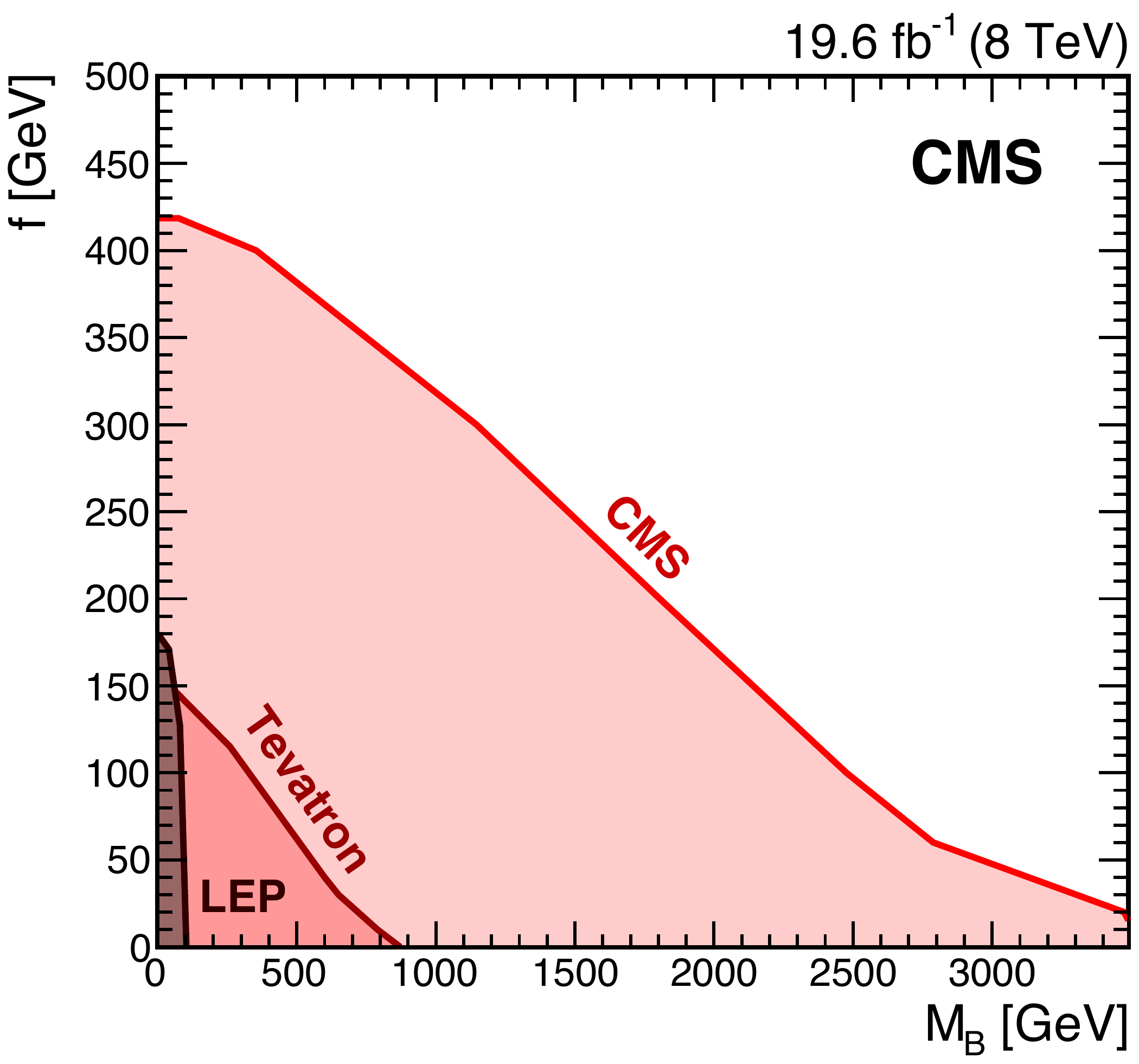}
  \end{center}
\caption{Observed 95\% CL lower limit on the branon mass $M_D$ as a function of the brane tension parameter $f$, together with the previous exclusion from the L3 experiment.\protect\cite{branons-L3} From Ref. \protect{\citen{EXO-12-047}}.}
\label{fig:branons}
\end{figure}

The monophoton analysis~\cite{EXO-12-047} also set limits on branons,\cite{branons,branons2,branons3,branons4,branons5,branons6} which are excitations of branes in models with extra dimensions inspired by string theory. These branons, just like string excitations, may be quantum precursor for black hole formation. Limits on branons as a function of the excitation mass $M_D$ and the brane tension $f$ are shown in Fig.~\ref{fig:branons}. This is the first dedicated search for branons since the LEP era.\cite{branons-L3} (The Tevatron limits shown in Fig.~\ref{fig:branons} are from phenomenological interpretation\cite{branons7} of the Tevatron monophoton and monojet data.)

\begin{figure}[hbt]
  \begin{center}
    \includegraphics[width=0.7\textwidth]{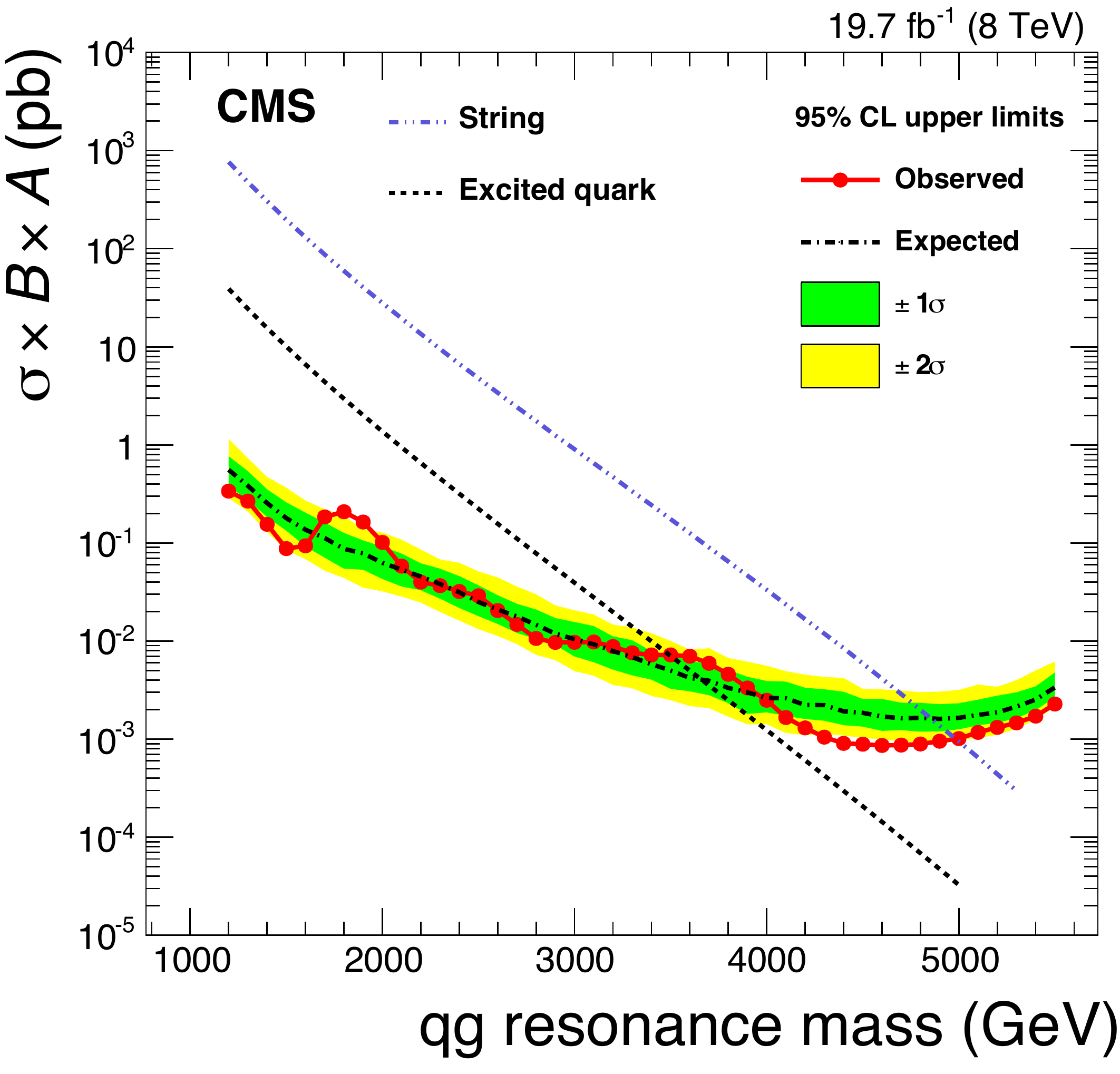}
  \end{center}
\caption{Observed 95\% CL upper limit on production cross section of quark-gluon dijet resonances (points) from the inclusive dijet analysis compared to the expected limits (dot-dashed) and their variation at $\pm 1\sigma$ and $\pm 2\sigma$ levels (shaded bands). Theoretical predictions for string resonance models are shown with the dash-dotted blue line. From Ref. \protect{\citen{CMSdijet8}}.}
\label{fig:strings}
\end{figure}

The dijet analyses also looked for string resonances, i.e., excited string states that mainly decay into $qg$ final states. These resonances are expected to be strongly produced and also mass-degenerate, with the mass close to the string scale $M_S$.\cite{strings,strings1} Therefore, the cross section for string resonance production is very large and stringent limits reaching 5.0 TeV are set on their existence (see Fig.~\ref{fig:strings}).\cite{CMSdijet8}

\begin{figure}[hbt]
  \begin{center}
    \includegraphics[width=0.7\textwidth]{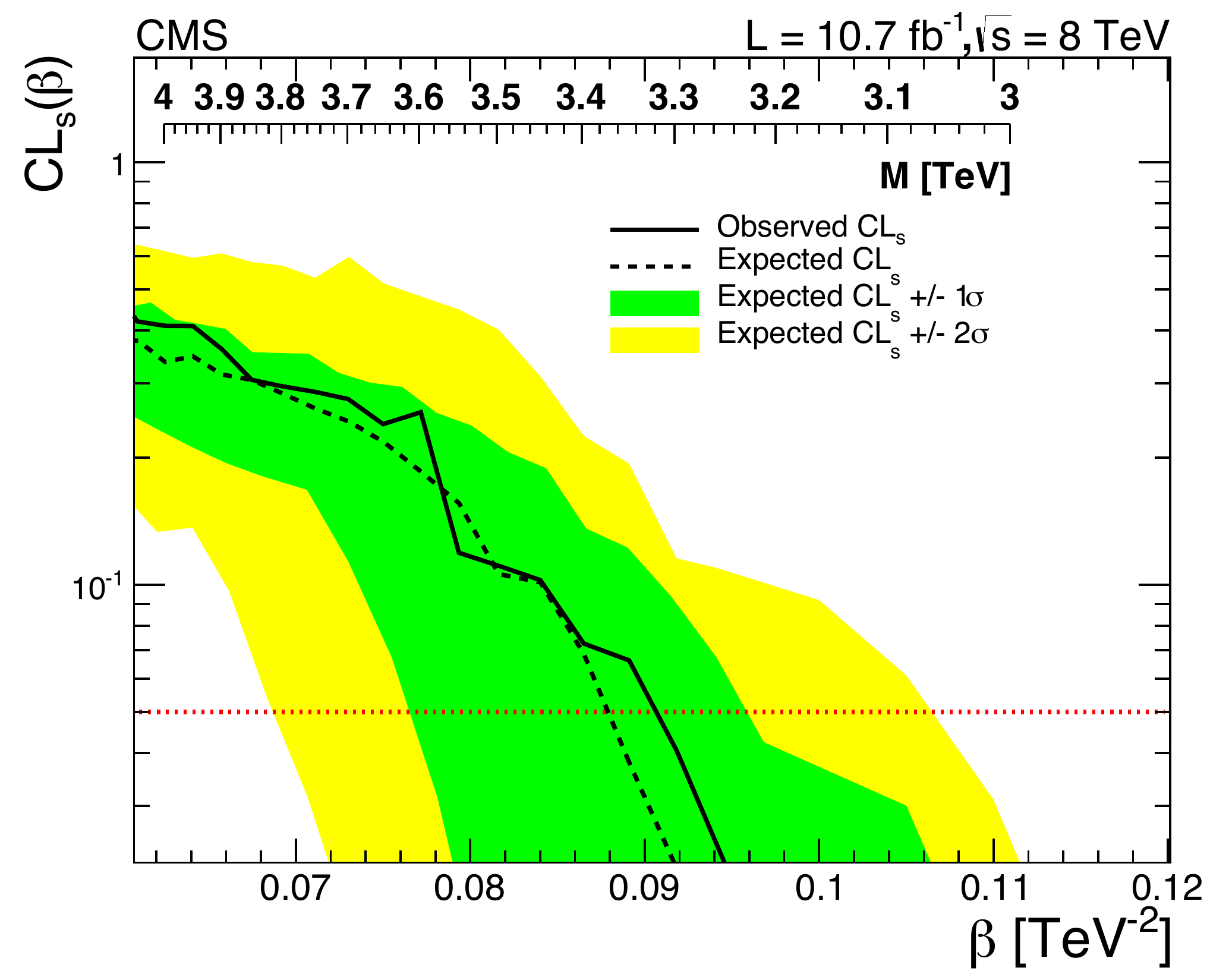}
  \end{center}
\caption{An observed (solid line) and expected (dashed line) 95\% upper limit on $\beta \equiv 1/M^2$, where $M$ is the extinction scale and its translation in a lower limit on $M$. The green and yellow bands indicate the $\pm 1\sigma$ and $\pm 2\sigma$ variations in the expected limit. From Ref. \protect{\citen{EXO-12-051}}.}
\label{fig:ext}
\end{figure}

Finally, there is also a search for jet extinction in the inclusive jet production, which is another way to looks for black holes or their quantum precursors. The idea is that jet production may be suppressed above a certain energy threshold where these new phenomena open up, which could be inferred from a deviation of an inclusive jet $p_T$ spectrum from a smooth power-law falling function.\cite{extinction} The CMS experiment has conducted the first search for this phenomenon to date using a data sample at $\sqrt{s} = 8$ TeV corresponding to an integrated luminosity of 11 fb$^{-1}$ and set a lower limit on the extinction scale of 3.3 TeV at a 95\% CL\cite{EXO-12-051}, see Fig.~\ref{fig:ext}.

\section{Conclusions}

A large variety of CMS analyses based on 2010--2012 LHC data resulted in stringent limits on the existence of extra dimensions and low-scale quantum gravity, in a number of models. In many cases, the limits obtained by CMS are the most restrictive to date. While the possibility to see these phenomena at the LHC is diminishing, the current searches are completely limited by the maximum machine energy of 8 TeV reached so far. There is still a lot of uncovered model parameter space left, which will be explored as early as this year by exploiting the LHC potential close to its design energy of 14 TeV.

\section*{Acknowledgments}

I'm indebted to my many colleagues, who built, commissioned, and ran the CMS detector, and produced all the beautiful results included in this review. These new results made working on the review an exciting and interesting task. This work is partially supported by the DOE Award No. DE-SC0010010-003376.

\end{document}